# Planetary Exploration Horizon 2061 – Report Chapter 5: Enabling technologies for planetary exploration

Manuel Grande[1], Linli Guo[2], Michel Blanc[3], Advenit Makaya[4], Sami Asmar[5], David Atkinson[5], Anne Bourdon[6], Pascal Chabert[6], Steve Chien[5], John Day[5], Alberto G. Fairén[7], Anthony Freeman[5], Antonio Genova[8], Alain Herique[9], Wlodek Kofman[9], Joseph Lazio[5], Olivier Mousis[10], Gian Gabriele Ori[11,12], Victor Parro[7], Robert Preston[5], Jose A Rodriguez-Manfredi[7], Veerle Sterken[13], Keith Stephenson[4], Joshua Vander Hook[5], J. Hunter Waite[14], Sonia Zine[9]

[1]*University of Aberystwyth, Aberystwyth, UK*

[2]*DFH Satellites co., CAST, Beijing, China*

[3]*Institut de Recherche en Astrophysique et Planétologie (IRAP), CNRS-University of Toulouse-CNES, Toulouse, France*

[4]*ESA-ESTEC, Noordwijk, The Netherlands*

[5]*Jet Propulsion Laboratory, California Institute of Technology, Pasadena, USA*

[6]*Laboratoire de Physique des Plasmas (LPP), Palaiseau, France*

[7]*Centro de Astrobiología (CSIC-INTA), Madrid, Spain*

[8]*Sapienza Università di Roma, Roma, Italy*

[9]*Univ. Grenoble Alpes, CNRS, CNES, IPAG, 38000 Grenoble, France*

[10]*Laboratoire d'Astrophysique de Marseille (CNRS- Aix- Marseille Université), Marseille, France*

[11]*Int'l Research School of Planetary Sciences, Universtà d'Annunzio, Pescara, Italy*

[12]*Ibn Battuta Centre, Université Cadi Ayyad, Marrakech Morocco*

[13]*ETH Zurich, Zurich, Switzerland*

[14]*Southwest Research Institute and University of Texas San Antonio, USA*

# Contents

*Summary:* The main objective of this chapter is to present an overview of the different areas of key technologies that will be needed to fly the technically most challenging of the representative missions identified in chapter 4 (the Pillar 2 Horizon 2061 report). It starts with a description of the future scientific instruments which will address the key questions of Horizon 2061 described in chapter 3 (the Pillar 1 Horizon 2061 report) and the new technologies that the next generations of space instruments will require (section 2). From there, the chapter follows the line of logical development and implementation of a planetary mission: section 3 describes some of the novel mission architectures that will be needed and how they will articulate interplanetary spacecraft and science platforms; section 4 summarizes the system-level technologies needed: power, propulsion, navigation, communication, advanced autonomy on board planetary spacecraft; section 5 describes the diversity of specialized science platforms that will be needed to survive, operate and return scientific data from the extreme environments that future missions will target; section 6 describes the new technology developments that will be needed for long-duration missions and semi-permanent settlements; finally, section 7 attempts to anticipate on the disruptive technologies that should emerge and progressively prevail in the decades to come to meet the long-term needs of future planetary missions.

# 1. Introduction

In this chapter we will consider the technological advances that we can expect and that will be useful to provide for the successful implementation of the challenging missions and investigations proposed in the previous chapter (chapter 4, report on pillar 2). The immediate task is to bridge the gap between past/present space missions and those for the next generation. We need to match up science requirements with available technologies.

A perfect historical example of the task is given by Galileo. He was interested in the scientific study of the Moon, so when he was shown a crude telescope used for maritime navigation, he had the inspired idea of pointing it at the Moon. Finding that he could indeed see previously unknown detail, he set about understanding and improving the crude device. This produced the first simple refracting telescope of the type we know today. He developed the theory so the construction could be understood systematically and developed. With the improved device he was able to make far more sophisticated observations of the Moon than had ever been achieved before, and incidentally to discover the moons of Jupiter, an event which many believe marks the foundation of modern astronomy. Moreover, the telescopes that were available to the world at large were now vastly superior.

Galileo followed the methodology which is core to our requirements for making step changes in planetary science capabilities:

- He Identified key scientific problems
- He took existing technology and developed it to a new level
- He then used it to solve problems, leading to a virtuous spiral of scientific and technical progress.
- He therefore left both science and engineering enhanced
- And he was DISRUPTIVE

It is likely that the future progress of planetary science on the 40-year timescale described in this book will be predicated on the successful identification, development, application and exploitation of new "disruptive technologies". These may come from within the planetary community or elsewhere. The current breakneck development of machine learning techniques is an obvious example. Our hope is that this chapter identifies many more.

But we must never forget that from our point of view, the science is primary. Understanding the correct instrumental approach requires us to understand the details of the required observations. This is what the previous chapters have done: chapter 2 explored the synergies between exoplanet and solar system science; chapter 3 then analysed six key science questions about planetary systems and identified the key observations that future planetary missions should perform to address these questions. These will be the major science drivers for the next 40 years. What methods will be needed to address these science challenges? Obvious categories are: autonomy, miniaturisation, power generation, instrument resolution, hostile environments, readout and communication speed and propulsion.

While it is clear that the requirements will drive improvements in science instrumentation over time, one must identify now the priority technologies that one must invest in immediately to meet these future science challenges, for example materials technology or computing. It is however also the case that the market will do a lot of this for us, and there is an obvious possibility for wasting resources on second guessing what will happen in the next 40 years. One must identify needed nearer term technologies that we can significantly improve, and how. And one must certainly be careful not to dilute resources by investing in technologies that look interesting but for which there are no clear-cut requirements in the planetary community.

In all this of course, one must not divert effort from making optimal use of data from past and present space missions by the broadest possible science community, in order not to waste good data and to correctly formulate new questions.

In the preparation of future planetary missions, one must remember that there is a hierarchy of exploration modes, which lends some predictability to the requirements. Earth-based observations and laboratory studies may be critical in order to correctly formulate the observational requirements of space missions proper. For example, one might need to understand both the laboratory spectrum of a particular mineral and its expression in a planetary environment.

In general exploration will progress in the following steps:

- Theory and modelling
- Remote sensing from Earth (surface or orbit), from Moon or L-point
- Interplanetary or interstellar (!) cruise
- Planetary flyby
- Planetary or Satellite Orbiter:
- Single platform
- Multi-platform/probes/penetrators
- SmallSats, nanosats and in future chipsats.
- Lander on planet or satellite surface, probes including balloons etc.
- Surface sample return + sample curation/analysis at Earth

One instrument development tool which has developed enormously, and indeed transformationally, over the last few years, is the detailed numerical simulation of a future instrument, not just in its measurement configuration, but in terms of its response to. a hostile environment, for example radiation or thermal and how these effects will impact not just its survival, but also its detailed measurement responses and capabilities. Instruments prototypes are now far more sophisticated than a couple of decades ago, and this progress will undoubtedly continue. It is particularly important that standards are further developed for these computational tools, so that design, ideas and processing can be shared among a team, and computing loads distributed. There is scope for far more detailed optimisation both in terms of measurement fidelity and resource minimisation than in the past. Indeed, minimisation of mass will be key to many of our most ambitious planetary explorations.

Of particular interest will be the development of surface sample return capabilities, and the associated curation, analysis and planetary protection. Meteorite studies can teach us much, as they can be analysed with highly capable equipment. Indeed, the Apollo program is a good example of where advanced analytical equipment developed for scarce Lunar samples has led to a revolution in terrestrial lab-based analysis with impact across a wide range of terrestrial applications. Yet meteorite studies do not give us any clue as to the geological context or age of the sample. For this, sample return or in-situ analysis is essential. Alteration, whether by meteorite ejection, space residence or spacecraft transfer will always be a problem. And there are many locations in the solar system from which we will never obtain an unambiguous meteorite sample. New discoveries from in-situ and remote studies will always raise questions that require complementary analyses, and preparation techniques in terrestrial laboratories allow for analyses at higher spatial resolution and precision. Moreover, analytical instrumentation and techniques on Earth will always have greater capabilities than their coeval in-situ counterparts.

However, the capabilities of in situ analysis continue to advance at pace, and the penalties of sample return missions in terms of complexity, weight, curation, avoidance of contamination and planetary protection will always be far greater than for in-situ analysis. So, it is likely that the future will still see a mix of in-situ analysis and sample return, with sample return reserved only for those locations where a detailed and complex analysis beyond the capabilities of ever evolving in situ studies can yield transformational results.

Whether for in –situ analysis, or sample return, we will clearly need to develop highly autonomous long-range surface, liquid or atmospheric mobility as well as sub-surface drilling capabilities. These will assuredly be complemented by highly effective autonomous navigation and in-situ sample selection, handling, preparation and analysis.

Human exploration, even by geophysical or astrobiological experts, will probably have a very low profile over the 40 years under consideration in this investigation, with the Moon and perhaps Mars as possibly the only exceptions.

Section 2 describes the future scientific instruments which will address the key questions of Horizon 2061 identified in chapter 3 (the Pillar 1 Horizon 2061 report) and the new technologies that the next generations of space instruments will require. From there, the chapter follows the line of logical development and implementation of a planetary mission, using the set of representative future missions identified in chapter 3 as a reference for the technology requirements to be met in the future: section 3 describes some of the novel mission architectures that will be needed and how they will combine interplanetary spacecraft and science platforms; section 4 summarizes the system-level technologies needed: power, propulsion, navigation, communication, advanced autonomy on board planetary spacecraft; section 5 describes the diversity of specialized science platforms that will be needed to survive, operate and return scientific data from the extreme environments that future missions will target; section 6 describes the new technology developments that will be needed for long-duration missions and semi-permanent settlements; finally, section 7 attempts to anticipate on the disruptive technologies that should emerge and progressively prevail in the decades to come to meet the long-term needs of future planetary missions.

# 2. Advanced instrumentation for the future

## 2.1. Introduction

In this section we aim to provide an overview or possible trends in future of space and planetary exploration instrumentation. The life cycle of a progression from remote sensing to in-situ observation to sample return will continue. We are already at the stage where our own Solar System has been remote sensed to a considerable degree. The need for further development depends on the nature of the science. Cartography will always be done with remote sensing, albeit strengthened with in-situ ground truth measurements, and eventual sample return. Geophysics is more reliant on improvements in in-situ capabilities. Atmospheric and plasma measurements can actually benefit from a return to more accurate remote sensing, in order to expand our current local measurements into a global understanding. The search for life, and also for a detailed isotope understanding leading into a comprehensive chronology and understanding, may require sample return. One of the main dynamics will be the trade-off between an ever evolving in-situ instrumental capability, and the need to perform analyses which can only be carried out in a terrestrial laboratory.

We will see continuous development, based on spin-offs from technology developments for terrestrial applications. An obvious example is the development, driven by Covid-19, of far more sophisticated in situ testing of biological samples. Other application may include printable instruments and electronics, cryoelectronics and thermoelectronics. A major driver will be the development of drones and smallsats for near Earth use. Many of these will probably be deployed as scouts for larger planetary missions. Another spin-off is likely to be derived from the enormous investment going into autonomous systems including autonomous cars. This is perhaps a classic example of an area where commercial development is attracting far more finance and research than space missions can, and we should be watching rather than funding current developments.

We would greatly benefit from new propulsion systems and technologies. For outer Solar System missions and for operations at the surface of planetary bodies, new methods for power generation particularly in the area of novel solutions to deploying nuclear power will be essential (see section C-3). Indeed, there is little point in pursuing other developments if a clear roadmap to implementation of future power requirements is not in place. The need for miniaturisation will be a key driver, and will lead to more and more emphasis on development of single chip instruments, of increasing sophistication.

The list of new instrumentation is by no means comprehensive. The different sections of this chapter are illustrative of useful trends and innovations, rather than a complete list. A good example is magnetospheric science and instrumentation. Most of the plasma instrumentation which has flown in planetary missions over the last 50 years would not be novel to our colleagues from 1970. What has taken place is a continuous evolution and increase in sophistication. There has been an evolution particularly in miniaturisation, and in onboard processing capability. The main innovations have actually been in remote sensing of plasmas, by neutral particle imaging (*refs*), and by XUV imaging (*ref* to SMILE). Perhaps the most important innovation has been the improvement in modelling made possible by vastly superior computing power, and the use of in situ data to provide adaptive input and validation to these models. In some ways, the evolution of planetary plasma instrumentation has been the opposite of that in most fields. Plasma instrumentation began with in situ measurements, which were subsequently incorporated into synoptic imaging, and used to provide ground truth. While it is clear that these trends will continue, particularly in producing highly miniaturised and sophisticated instruments for constellations of small probes, we have not identified new measurement approaches which at this stage need to be called out in a section of the book.

**Remote sensing instruments**

A range of instruments will be developed to provide higher definition spectral, spatial and temporal resolution, with improved SNR (Signal over Noise Ratio), autonomous interpretation. New detectors will be deployed from new vantage points. In all cases, low mass will be a key driver.

**In situ instruments:**

One theme which we know will drive step changes in in-situ instrumentation will be the drive to detect signatures of past and extant life. This is perhaps the major growth area. New tools and methods will certainly include the ability to investigate biosignatures at nanoscales, using analytic techniques, based on analytic electron microscopy, mass spectroscopy, and surface and isotopic analysis.

Other important strands will include the further development of better drilling technologies, in a variety of terrestrial and outer planet moon environments.

These investigations will be facilitated by advances in autonomy and automated systems, both for sample acquisition and sophisticated analysis, probably facilitated by machine learning and AI derived expert systems. This will certainly remain true for those parts of the Solar System which are more difficult to access. In fact, for the forty-year period here considered, it is likely that only Mars and asteroids will be a target for sample return, although preparatory work for other locations will begin. Currently, a sample return mission to Ceres (reference to GAUSS mission) is amongst the most likely.

Our scientific objectives in terms of detection of precursors to DNA or analogous processes in the search for traces of life will certainly evolve, as will our understanding of the information that can be derived from isotopic signatures. These more sophisticated objectives will of course drive more sophisticated instrumentation, and the answers we will thus obtain will in turn drive a next generation of more sophisticated science questions.

A need to operate in more hostile environments, in terms of radiation, pressure, extreme temperature (hot or cold) and corrosion, will require development of appropriate instrumentation. For these cases, we will likely not be able to rely on repurposing terrestrial instrumentation, and will need to develop specific solutions. This will certainly require foresight, and considerable investments in time and money.

There will be a need for long range, autonomous rovers, which do not need continuous and expensive supervision.

We will certainly see improvements in drilling, coring and sampling techniques for varied terrains and environments, and new seismometer arrays. Delivery by penetrators, and multipoint measurements will also require easily calibrated, rugged instrumentation, to go with the novel delivery systems. A possible timeline for in situ sample analysis instrumentation might encompass:

- Near future: Provide elemental, molecular, isotopic information at scales down to a single atom; preserve organics and hydrated minerals
- Medium term: Cryogenic (100K) vacuum transfer techniques for of water ices and volatile organics
- Long term: be able to do everything at 10K or 800K

**Sample return**

Sample return requires the development of a range of new techniques, in addition to those outlined above. In the next forty years, these will probably be required for sample return from the Moon, Mars and asteroids. Each will provide particular challenges for sample selection, retrieval and handling, encapsulation and return.

Among the required techniques will be:
- Scooping and digging
- Sampling on slopes
- Melting subsurface ice-
- Cryogenic sample return capabilities

For the farther future, there will be preparations for
- Cryogenic operation
- Hot sample return
- Contamination control

All these must be complemented by developments of advanced Earth-based laboratory analytical tools and models. Development of curation facilities on a timescale appropriate to missions in active development will also be key.

There may however be, post Covid-19, an increased reluctance to return potentially bioactive samples to Earth.

**Disruptive technologies**

Finally, it is clear that a great deal of progress can, and indeed must, be driven by so called "Disruptive Technologies. Machine autonomy will be key in many areas, such as navigation, sample collection, and pattern recognition for data compression. New machine learning algorithms, carried over from terrestrial developments, will be behind many of these innovations.

Progress in cloud computing, quantum computation, big data, modelling and simulation, will be extremely rapid, if perhaps unpredictable. Geographic information systems for example, have already transformed the ways we think about using cartographical data. One key requirement is the implementation of international and multi-agency standard to ensure interoperability.

As remarked above, the areas of current rapid technological progress are in biosciences, machine learning and fabrication. It is these breakthroughs which we must continue to monitor, and to incorporate and develop in our future planetary investigations, just as Galileo did!

## 2.2. Advanced sensors and scientific investigations for the characterization of planetary environments, surfaces and interiors

### 2.2.1. Instruments for Planetary gravimetry and geodesy

Geodetic and geophysical investigations have enabled outstanding scientific discoveries regarding the internal structures of planets (Margot et al, 2007), moons (Schubert et al, 2007), asteroids (Cheng et al, 2004), and comets (Pätzold et al, 2007). An accurate knowledge of celestial bodies' interior is fundamental to enhance our understanding of the formation and evolution of the Solar System. Gravimetry and radio science experiments have played a crucial role in space missions devoted to survey interplanetary bodies by providing measurements of their gravity fields. The scientific objectives of gravity investigations have been achieved through the analysis and processing of spacecraft radio tracking data collected by networks of ground stations (e.g., NASA Deep Space Network, ESA European Space Tracking network). Spacecraft have been equipped with onboard radio science systems to establish telecommunication with the ground antennas that enables the acquisition of radio tracking data. The analysis of these observations, which inform on the relative distance and velocity between the probe and the Earth's station, allows measuring indirectly the gravity field of the central body. An accurate estimation of the gravitational anomalies is fundamental to fully characterizing the internal mass distribution of the celestial body.

The architecture of the radio science instrumentation is generally based on elements of Telemetry, Tracking & Command (TT&C) systems. The accuracy of the radio tracking data for deep space navigation is limited by the presence of Earth's ionosphere and troposphere, thermal, and, especially, solar plasma noise. To improve the performances of the radio science systems, a dedicated transponder has been included in new radio science systems. This novel scheme yields a full calibration of the plasma noise with a three-link configuration (i.e., X/X, X/Ka, and Ka/Ka). The NASA mission Cassini demonstrated the benefits of this calibration technique (Kliore et al, 2007), and the dual-transponder architecture has also been adopted for the NASA mission Juno (Asmar et al, 2017), and the ESA missions BepiColombo (Less et al, 2009) and JUpiter Icy moons Explorer (JUICE) (Grasset et al, 2013). The radio tracking data accuracies that are achievable with this instrument are ~3 $\mu m\, s^{-1}$ at 1000-s integration time and ~20 cm for Doppler and range, respectively (Less et al, 2014).

Multifrequency radio systems solved the major weaknesses of single-band radio systems by significantly reducing the data noise and by enabling data acquisition in proximity of superior solar conjunctions. Future exploration of planets and moons, however, will require more demanding measurement precisions for interdisciplinary investigations that aim at addressing fundamental scientific questions regarding the formation and evolution of planetary systems.

An alternative radio science configuration that would fulfil these more challenging requirements is based on inter-satellite tracking between a pair of spacecrafts. The missions Gravity Recovery and Climate Experiment (GRACE) (Tapley et al, 2004) and Gravity Recovery and Interior Laboratory (GRAIL)(Zuber et al, 2013) demonstrated the benefits of measuring orbital changes along the line connecting two satellites and, in particular, the relative velocity by means of the inter-satellite radio tracking data Doppler shift (Wolff et al, 1969) . GRACE and GRAIL inter-satellite observations were ~2-3 orders of magnitude more accurate than deep space tracking data leading to extremely precise determination of the gravity fields of the Earth (Tapley et al, 2005) and the Moon (Zuber et al, 2013), respectively. An accurate high-resolution mapping of celestial bodies' gravitational anomalies will require extremely precise radio tracking data to reveal the properties of their internal structures.

This level of precision of gravity measurements can be also obtained by direct measurements of the gravity forces through a gradiometer. The ESA mission Gravity field and steady-state Ocean Circulation Explorer (GOCE) (Muzi et al, 2004) was designed to host onboard three pairs of accelerometers that measured gravitational gradients along three orthogonal axes. The analysis of these data allowed mapping accurately the ocean currents (Knudsen et al, 2011) and revealing details of the geological structure of the Antarctic continent (Gilardoni et al, 2011).

GRACE-like and GOCE-like missions to planets and moons would exclusively include a science payload devoted to gravity investigation. These technologies require sophisticated elements with high mass and power demand. Space missions that will be designed to study poorly explored worlds in the Solar System will be focused on multidisciplinary investigations. Therefore, the development of a novel instrumentation that enables gravity experiments with the level of accuracy of GRACE/GRAIL and GOCE data, and the possibility to include other science instruments onboard will be crucial for the future exploration of the Solar System.

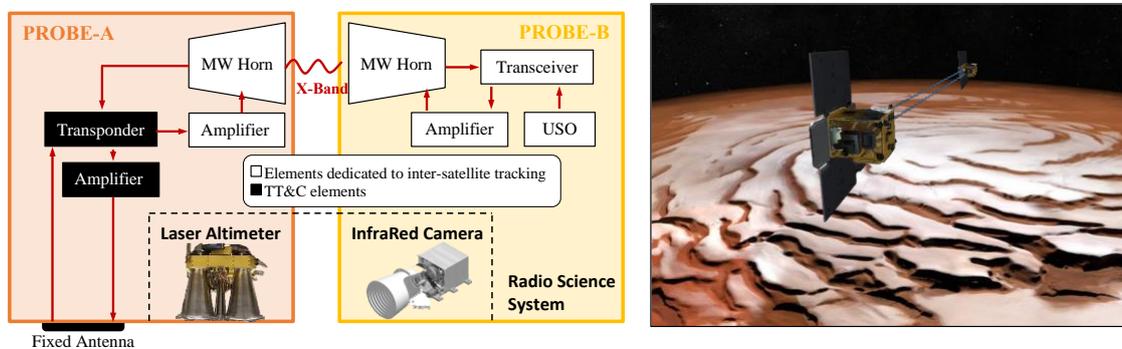

*Figure 5.1 Block diagram of a dual-spacecraft science payload that includes: (1) a novel radio science system; (2) a laser altimeter based on the MLA (credits: NASA GSFC); (3) Infrared Camera based on TIR-S onboard Hayabusa-2 (credits: JAXA). The right panel shows a mission concept of the dual-spacecraft configuration to study the Martian climate.*

A novel inter-satellite tracking system in line with science and engineering constraints is based on the solid heritage of existing technologies developed for deep space tracking. This radio science architecture shares part of TT&C elements with the inter-satellite tracking system including a transponder (e.g., Integrated Radio science and Deep Space TT&C Transponder (Arena et al, 2016). One of the satellites hosts an Ultra-Stable Oscillator (USO) and a transceiver to transmit a signal to the other satellite enabling extremely highly accurate two-way inter-satellite Doppler data. This radio system configuration also provides deep space radio tracking data. A combined analysis of inter-satellite and deep space Doppler data is fundamental to determining short- and long-wavelength gravity anomalies that provide crucial constraints on the celestial body's outer layers and deep interior, respectively.

The compact design of this radio science system will be well-suited for large-scale missions with multiple science instruments, and for small- and medium-class missions dedicated to outstanding geodetic and geophysical investigations. A comprehensive understanding of the geophysical properties of planets and moons is achievable with a dual-spacecraft mission with this radio science system, a laser altimeter (e.g., Mercury Laser Altimeter, MLA (Cavanaugh et al, 2007) and a Thermal Infrared (TIR) camera (e.g., TIR-S) (Okada et al, 2017). Figure 5.1 shows the block diagram of this science payload onboard two small satellites that orbit Mars to thoroughly investigate its climate evolution (Genova et al, 2020).

## 2.2.2. Radar instruments

Radars are active instruments: an electromagnetic wave is transmitted from the antenna towards the observed surface, part of the signal is reflected by the surface, while another part propagates in the subsurface, to be then reflected by and/or transmitted through any interfaces. The wave is affected by scattering at interfaces, as well as by scattering and absorption in the medium.

High-frequency imaging radars have been in use for 50 years in Earth Observation: their value for planetary science has been demonstrated from Magellan (Saunders et al, 1992) to Cassini (Lopes et al, 2019), to map planetary and satellite surfaces. The main evolution in the last decades is that SAR instruments for planetary surfaces have benefitted from technologies developed for Earth Observation like VenSAR (under study for EnVision, ESA M5 mission), a phased array synthetic aperture radar inheriting from NovaSAR-S developments (Ghail et al, 2018). For the next instrument generations, the key instrumental developments are expected to come from Earth Observation, so these perspectives are not be developed in this section: we concentrate on sounding radars for planetary subsurface, and on low-resource instruments for small bodies.

Spaceborne sounding radars have been flying on planetary missions for nearly three decades and are becoming classical tools to probe the subsurface and internal structure of solar system bodies: MARSIS onboard Mars Express (ESA) (Picardi et al, 2005) and SHARAD onboard MRO (NASA) (Seu et al, 2007) have imaged Martian subsurface structures, especially within the ice caps (Figure 5.2, right panel), offering constraints on composition. More than 30 years after ALSE onboard Apollo 17 (NASA), LRS on board Selene-Kaguya (JAXA) (Ono et al, 2009) probed the Moon regolith. MoSIR onboard Tianwen-1 (CNSA) is arriving in February 2021 on Mars (Zou et al, 2021). Finally, RIME (Figure 5.2, left panel. Bruzzone et al, 2015) and REASON (Phillips and Pappalardo, 2014) are under implementation for the JUICE (ESA) and EUROPA Clipper (NASA) missions to the Jovian satellites, and a sounding radar instrument under study for EnVision (Bruzzone et al, 2020). Such nadir-looking sounding radars are particularly sensitive to surface roughness. A large roughness with respect to the wavelength results in lower penetration depths, the radar operating then as an altimeter, while a low roughness allows dielectric mapping and detection of internal reflections, provided that some surface elevation model is available.

Scientific returns of such missions show radars as a unique opportunity to access the third dimension, by providing a direct measurement of the bodies' interior thanks to kilometers-deep penetration, and providing context for remote measurements of surfaces and stratigraphic connection of the observed terrain units. Nevertheless, radar instrument design remains challenging: instrument performances in terms of investigation depth, sensitivity and resolution are highly dependent on the considered wave frequency band - and on the composition and structure of the fathomed bodies, which are generally unknown. Performances are also strongly dependent on the geometry of observation: incidence angles, measurement orbit arcs, multi-sensor geometry… Therefore, instruments have to be significantly revisited for each mission with major trade-offs related to antenna accommodation, data flow and potentially available power.

With SHARAD, RIME and REASON, a true industrial subsector in planetary radar has emerged, comparable to what exists for Earth observation instruments, and instrument concepts similar to those planetary radars are proposed in order to study Earth's polar caps. Future instruments will benefit from increased electronics integration and induced mass reduction.

After the nadir-looking configuration, next generation instruments would consist in slant-looking P-Band polarimetric radars (Campbell et al, 2004), benefiting from increased telemetry system performances to image the geological structure of the first tens of meters. For Mars, imaging the near subsurface is crucial to model geophysical processes for both geological and exobiology purposes, while mapping the near-surface ice is necessary for future human exploration of the Moon and Mars. These instruments will inherit from Earth Observation missions' developments, especially large antenna structures such as the 12 m-wide deployable reflector developed for the BIOMASS / ESA mission (Le Toan et al, 2011).

The instruments developed for planets and large satellites observation are resource-demanding: large antennas, high power, huge data volume (Figure 5.2, left panel). Generally, the whole mission is designed around the instrument. Such instruments and missions would benefit more and more from developments made for Earth Observation, but also from increased data rates, especially concerning deep space missions.

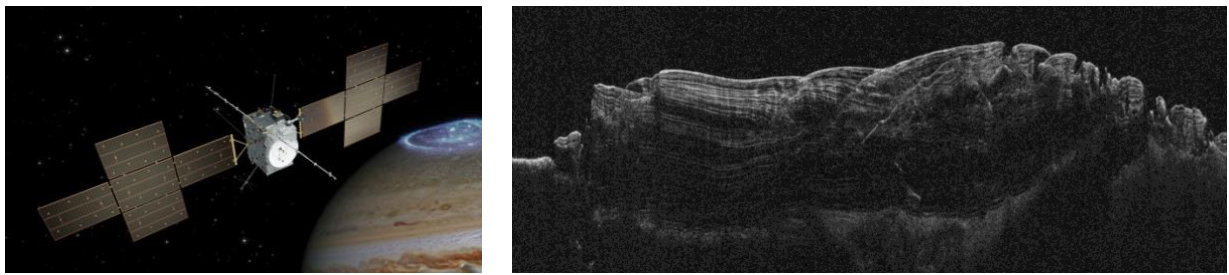

*Figure 5.2. Left panel: Artist's impression of the JUICE spacecraft, showing the size of the RIME radar antenna (© ESA). Right panel: A SHARAD radargram extracted from the PDS (Campbell and Phillips, 2014), displaying layering in Mars' North Polar Deposit (orbit 2612).*

On the other hand, less demanding instruments can either be part of a large payload, either be embedded on a small CubeSat-like platform. For both planets and small bodies, small platforms or CubeSats are also an opportunity to develop bistatic or multi-sensor measurements and satellite formation flying, and to increase sensitivity and performances.

Radars for icy and rocky small bodies require specific developments. Orbit configuration, relative speed, altitudes are so different for planets and small bodies that radar instrument concepts and designs deviate significantly from orbital planetary radars (Herique et al, 2018). Thus orbital radars dedicated to small bodies require a large versatility in term of operation range and observation angles, made possible by a low relative speed (lower than few tens of meters per second for a body size lower than ~10km). For kilometric-sized bodies or smaller, bistatic radars can operate in transmission or in reflection between a lander and one or two orbiters (Figure 5.3). This original concept provides a direct measurement of the average permittivity of the body and of its internal structure and stratigraphy: CONSERT onboard Rosetta (Kofman et al, 2015) fathomed a limited part of the nucleus of comet 67P/ Churyumov–Gerasimenko. The next generation of instruments will focus first on fathoming small bodies' regolith at higher resolution, to understand evolution processes, and also in probing the deep interior with lower frequency radars to better model accretion and re-accretion processes (Herique et al, 2018, 2019). JuRa is such a monostatic low frequency radar, under development for Juventas' CubeSat onboard HERA (ESA) mission, to fathom the internal structure of Dimorphos (Herique et al, 2020).

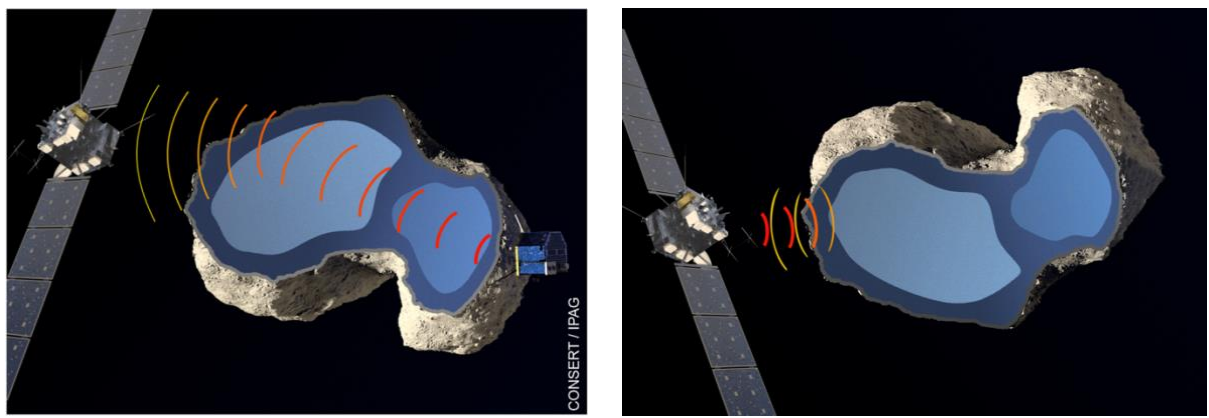

*Figure 5.3. Left panel: CONSERT-like configuration, with a bistatic radar operating in transmission between a lander and an orbiter. Right panel: JuRa-like monostatic configuration, probing the regolith and the interior. A*rtist view developed from CONSERT/Rosetta (credit: CGI/Remy Rogez; shape model: Mattias Malmer CC BY SA 3.0, Image source: ESA/Rosetta/NAVCAM, ESA/Rosetta/OSIRIS/MPS/UPD/LAM/IAA/SSO/INTA/UPM/DASP/ IDA).

New-generation small radars, especially when exploring comets and asteroids, require dedicated development to benefit from increased electronics integrations and the induced reduced mass and increased efficiency with hybrid and ASIC components envisaged.

### 2.2.3 Radio and Optical Link techniques
**The Future of Radio and Optical Link Science**

From Mercury to the outer reaches of the solar system, the past six decades have witnessed a vast set of discoveries utilizing radio science (RS) methods. For example, based on key evidence from gravity radio science, sub-surface oceans have been inferred at Titan, Enceladus, and Europa, where potential future missions may search for life (Asmar et al., 2020).

The ability to precisely measure properties of spacecraft radio signals — frequency, phase, delay, amplitude, polarization — provides unique leverage to extract new information about atmospheres, ionospheres, rings, surfaces, shapes, and internal structure of solar system bodies (Asmar et al., 2019).   In addition to planetary sciences, RS observables such as precision Doppler and ranging are critical to studies in fundamental physics and solar dynamics such as observing effects on signal passage through the Sun's gravitational field and solar wind, investigating gravitational waves, and monitoring planetary motion to study gravitational theory and solar mass loss (Armstrong et al., 2003; Genova et al., 2018; Smith et al., 2018; and Woo, 1993; Armstrong, 2006).

RS remains a powerful and cost-effective tool for many solar system investigations planned or conceived in the coming decades. Additional science discoveries could be enabled by developing new technologies and mission concepts such as:

- Deployment of small spacecraft, especially in orbiting constellations for high spatial and temporal resolution of atmospheric and gravitational mapping (see illustration),

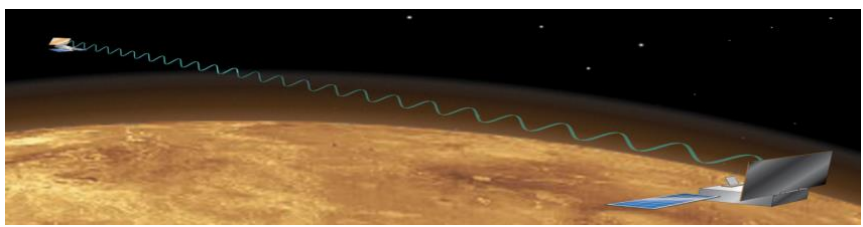

*Figure 5.4. Illustration of a two-member constellation of small spacecraft at Venus with crosslinks for radio occultations that can lead to global coverage with high spatial and temporal resolutions.*

- Novel instrumentation and calibration techniques to improve data quality by up to an order of magnitude over current levels,
- Exploitation of uplink transmissions from Earth such as those used by New Horizons at Pluto, to improve sensitivity by orders of magnitude, and
- Development of optical link capabilities for new high-precision link science using future optical communication systems.

**Future Mission and Experiment Concepts**

In the next decade, technological advances, and use of small spacecraft with improved radio capabilities that approximate those of much larger spacecraft could enable many additional scientific breakthroughs in atmospheric dynamics, interior structures, and surface properties. Development of new techniques and technologies to reach these goals can enable all solar system missions to benefit from this enhanced low-cost science capability.

***Spacecraft Constellations for Atmospheric Structure & Dynamics****:* Compared to spacecraft-to-Earth radio occultations (RO), substantial increases in global coverage as well as spatial and temporal resolution are possible using crosslinks among a constellation of small spacecraft, akin to GPS-based RO constellations at Earth. At Mars and Venus, for example, simulations show that nearly global coverage can be accomplished in a couple of weeks with two spacecraft in optimized orbits as compared to many years via the traditional downlink to Earth. Such crosslinks (see crosslink demo carried out with Mars orbiters in Ao et al., 2015) can profile the atmosphere through clouds and aerosols and detect the signature of phenomena such as gravity waves, planetary waves, turbulence, and jet streams, and for Venus, the possibility of detection of seismically-induced waves, potentially opening a window into possible current tectonic activity.

***Spacecraft Constellations for Interior Structure & Dynamics***: Precision gravity experiments via spacecraft-to-spacecraft crosslinks have been utilized at Earth (GRACE, Tapley et al., 2004) and the Moon (GRAIL, Zuber et al., 2013), and SELENE's farside subsatellite for high-impact studies of planetary interiors and monitoring mass transport. Applying this concept to other rocky, icy, or gaseous bodies, would enable rapid high-resolution global coverage and provide stringent constraints on the thickness of ice and the characteristics of subsurface oceans at icy moons.

***Differential Interferometry for Planetary Rotational State & Tidal Deformations***: Tracking multiple landers or spacecraft from one or many Earth stations suppresses most common noise sources, enhancing the sensitivity to the desired science goals (e.g., gravitational fields, rotational states, tidal deformations, and winds). The benefits of same-beam differential interferometry were recognized by (Counselman et al.,1972) and advanced by (Preston, 1974; Edwards et al.,1992; Bender,1994 and others), and applied to examine librations (ALSEP) and gravity field (Selene) of the Moon, studying winds on Venus (Pioneer probes and Vega balloons), and observing Huygens' descent on Titan. New uses have been proposed for the Moon, Mars, and Jupiter.

***Doppler Wind Experiments for Measurement of Atmospheric Dynamics***: Radio links from probes descending through planetary atmospheres provide information on atmospheric dynamics as demonstrated at Venus, Jupiter, and Titan. Links can be established with proximity spacecraft or ground stations, and, if multiple links are possible, recovery of complete wind vectors can be made. In some cases, ranging, VLBI techniques, and onboard observations provide additional information. A constellation of small probes and/or balloons can potentially provide a global assessment of atmospheric dynamics. Absorption on the probe's radio link can be used to infer the integrated abundance of ammonia (giant planets), sulfuric acid (Venus), or other absorbers. Similar techniques can be applied to atmospheric balloons as was done on the VEGA mission at Venus.

***Scattering Studies for Surface Properties***: In a bistatic scattering experiment, a spacecraft downlink signal is reflected off a planetary surface and received on Earth or another spacecraft. Information can be deduced about the surface and near subsurface electrical properties and roughness. Reversing the path via an uplink provides higher signal-to-noise ratio (SNR), as shown by New Horizons and Lunar Reconnaissance Orbiter (LRO); multiple simultaneous uplinks with frequency offset can improve the SNR and resolve subsurface degeneracies. A constellation of small spacecraft makes global characterization feasible and yields information at scales useful to the safety of planned landers. Scattering from subsurface structures provide added exploration opportunities where redundancies can be resolved using polarization, wavelength, and obliquity variations. Use of bistatic techniques with landed spacecraft can provide finer spatial resolution.

| Investigation | Representative Science Goals | Technology Goals |
|---|---|---|
| **Interiors** | • Core & mantle size and density <br> • Interior structure models <br> • Formation mechanism <br> • Thermal history <br> • Tidal and natural oscillations | • Spacecraft constellations crosslinks <br> • Same antenna beam tracking <br> • Advanced radiometric calibrations <br> • Atomic clocks |

| Investigation | Representative Science Goals | Technology Goals |
|---|---|---|
| | | • Small spacecraft orbit insertion |
| **Atmospheres, Rings, & Other Media** | • Temperature-pressure profiles<br>• Neutral & ionospheric densities<br>• Wave structures and origins<br>• Shape models | • Spacecraft constellations crosslinks<br>• Advanced software-defined radios<br>• Miniaturized stable oscillators<br>• Small spacecraft orbit insertion |
| **Atmospheric Winds** | • Wind velocities | • Small spacecraft delivery/entry |
| **Surfaces** | • Electrical properties & roughness | • Uplink & crosslink instrumentation |
| **Surface-Atmosphere Interactions** | • Seasonal ice deposition on Mars<br>• Geophysically-induced waves | • Geographical & temporal resolution |

*Table 5.1: Future Radio-Optical-Link Science and Technology Goals*

**Current Status and Future Capabilities Needed**

***Small Spacecraft***: The highest technology priority is the maturation of multi-band crosslinks between robust small spacecraft in suitable planetary orbits. This requires advances in miniaturized radios, ultra-stable oscillators (USO), and orbit insertion. A critical need is the development of key technologies for software-defined radios (SDR) size-suitable for small spacecraft. The Iris SDR used for the MarCO mission provided experience for further advances towards science-quality radiometric and functional capabilities. USOs have always been critical instrumentation for radio occultation experiments. Small spacecraft cannot accommodate the mass and power of the current top-of-the-line flight USO (Allan deviation $\sim 2\times 10^{-13}$ at ~100 s), so, the goal is to mature smaller units with comparable stability based on newer technologies.

***Advanced Link Calibrations:*** A leading source of radiometric noise is Earth's troposphere, which is partly calibrated via water vapor radiometers. An order of magnitude precision improvement is possible by pointing along the same path as the signal arriving at the antenna, as opposed to the current generation that resides adjacent to the antenna. The next noise source is unmodeled mechanical motion of the ground antenna's phase center. A method for suppressing this intrinsic noise is to combine the 2-way data from the principal antenna with Doppler data received at a smaller stiffer antenna.

***Enhanced Science via Atomic Clocks:*** Future one-way or crosslink observations enabled by onboard atomic onboard clocks with improved stability and reduction in size, could achieve performance comparable to two-way coherent links referenced to ground station atomic clocks. This enables nearly continuous tracking via smaller ground stations, enhancing the science at lower cost. A miniaturized prototype Mercury-trapped-ion atomic clock (DSAC-1) with Allan deviation of ~$2\times10^{-14}$ (1000s) is being tested in Earth orbit and DSAC-2 is offered to Discovery missions as a technology infusion. An even smaller version with comparable stability is under development.

***Enhanced Science via Optical Links***: Optical links are expected to be demonstrated on future deep space missions. Laser links can provide measurements in most areas already pioneered by RS but, in many cases, with new scientific capabilities and precision. Occultations could better probe upper atmospheres and Doppler measurements could provide improved gravity fields measurements. In the next decade it is important to lay the technological framework for future optical link science.

## Recommendations for Coming Decades

Scientific studies using spacecraft radio links have led to numerous discoveries. In the next decade, technological advances, and use of small spacecraft with increasing radio capabilities that approximate those of full-scale spacecraft, could enable many additional scientific breakthroughs in atmospheric dynamics, interior structures, and surface properties. Development of new techniques and technologies to reach these goals can enable all solar system missions to benefit from this enhanced low-cost science capability.

## 2.3. Principles and Instrumentation for in situ investigations of dust and gas

### 2.3.1. Dust investigations

**Cosmic dust measurements - major motivation and types of dust**

Cosmic dust can be divided in two categories: interstellar dust (ISD) — 'visible' as the dust blocking the light of the stars of the milky way — and interplanetary dust particles (IDP) — visible as the zodiacal light, which is sunlight, scattered by interplanetary dust particles. Interstellar dust particles reside in diffuse or in dense clouds and are the basic building blocks of what eventually become stars, planets and later on life. It plays a crucial role in astrochemistry, for cloud thermodynamics, and characterising ISD is important for astronomical observations: it is the medium we look through to observe the universe, and the dust physical properties are needed for e.g. interpreting observations of far away protoplanetary disks. Classical astronomical observations of ISD over long kpc scales are used to reveal ISD composition and size distribution using measurements of wavelength-dependent extinction and polarisation of starlight, emission by the dust in the infrared, and observations of chemical abundances in the gas (assuming the missing elements, in comparison with the abundances of a reference like the Sun, are locked in the dust). Using this ensemble of observations, models are built for ISD size distribution and composition.

In 1993, a new type of observations became available, providing ground truth information on ISD: for the first time, interstellar dust had been detected in situ in the solar system with a dust detector on-board of the spacecraft Ulysses. This is possible thanks to the the relative motion of the solar system and the Local Interstellar Cloud. Ulysses flew out of the ecliptic plane, and its orbit being almost perpendicular to the inflow direction of ISD has facilitated distinguishing interstellar from interplanetary dust. Ulysses has detected between about 500 and 900 particles during 16 years and opened the era of in situ ISD research in the solar system. More observations followed (Galileo, Helios, Cassini) and in 2016, the Cassini Cosmic Dust Analyser (CDA) measured the composition of 36 ISD particles, whereas the Stardust mission brought back some samples of ISD in its sample return capsule (2006, with analysis in 2014). A comprehensive review on interstellar dust in the solar system (incl. relevant references) is given in Sterken et al., 2019.

The zodiacal dust on its turn has been explored with in situ detectors already since more than half a century! Besides ordinary IDPs, also various types of dust 'between' the planets have been examined, for instance, dust coming from active moons, stream particles, planetary rings, cometary dust, and dust clouds around airless bodies.

Enceladus is an example of such an active moon with a subsurface ocean, where water ice particles escape via vents into space. Cassini CDA (a time-of-flight mass spectrometer) measured the composition of the dust particles in Enceladus' plumes and in Saturn's E-ring, illustrating that subsurface ocean compositions can be probed without the need for landing on such a moon. Also Io has volcanoes whose tiny particles get accelerated in the Jovian magnetic field and become very fast nanometer-sized 'stream particles'. Their composition was also measured by Cassini CDA. Dust impacts on airless bodies cause ejecta and as such, airless moons are surrounded by an ejecta cloud. Measuring these ejecta can be used to compositionally map the surfaces of these moons without a landing (Postberg et al, 2011).

Thus, the importance of in situ cosmic dust measurements goes far beyond "just" measuring dust. Besides this, the interplanetary dust — which mostly stems from comet activity, and asteroid collisions — provides us insights in the history of our solar system, these particles are also a means towards understanding geochemical conditions on subsurfaces of active moons, and towards probing the surface composition of airless bodies. Also physical processes of the dust and dust as charged probes to investigate the Interplanetary Magnetic Field (IMF) or planetary magnetospheres are subjects of study. A comprehensive review of interplanetary dust (incl. relevant references) is given in Grün et al., 2019 and Koschny et al., 2019.

**Dust measurement methods**

What methods are used for probing the dust in the solar system? A first category are **ground-based** heritage instruments: classical astronomical observations are made of the zodiacal cloud, comets and active asteroids, mostly in the optical part of the spectrum using ground-based telescopes. Millimetre-sized or tens of micrometer-sized dust particles that enter the atmosphere, become meteors that can be measured by **optical** (photo, video, fireball networks) and **radar instruments**. Two types of radars exist: the specular backscattering meteor radars (e.g. AMOR, CMOR) which measure the ionised-atmosphere tail of a meteor, and the high power large aperture (HPLA) radars like e.g. Arecibo (until 2020) measure *head echoes*. This is the portion of the atmosphere that ionises in front of the meteor; it has a shorter timespan and moves along with the meteor.

Smaller interplanetary dust particles of only one to a few micrometer in size that slowly move downwards through the atmosphere, have been collected in the stratosphere using **aeroplanes**. These planes have collectors with an oily substance, and the particles can be analysed in the laboratory. Also analysed in the laboratory, but on the large end of the scale-bar, are the **meteorites** that are found everywhere on Earth, but especially in the Arctic or Antarctic and in deserts they are easy to identify and collect.

**Space-based telescopes** observe dust in the infrared. For instance, the Spitzer space telescope measured several cometary trails, and IRAS and COBE satellite data were used for mapping the zodiacal cloud [Rowan-Robinson et al., 2013]. **Cameras** on-board spacecraft take detailed images of dust coming off from comets (e.g. Rosetta - OSIRIS). **Sample return missions** bring back samples of dust from cometary flyby's (e.g. Stardust) or asteroid rendez-vous (Hyabusa, OSIRIS-Rex), and from interplanetary space (e.g. Stardust InterStellar Preliminary Examination).

The workhorse of space-based dust instruments are the **impact ionisation instruments**, because of their high sensitivity (also to small dust particles: "nanodust" down to $10^{-21}$ kg) and their reliability (multi-coincidence detectors). Examples are the Ulysses Dust Analyser, Galileo Dust Detection System, and Cassini Cosmic Dust Analyser. **Time-of-Flight mass spectrometers** are a derivative of these types of instruments, and provide the elemental composition of the particles (e.g. Cassini CDA, Stardust CIDA). The working principle of these instruments is the following: upon impact of a dust particle on the target of such instruments, the particle and some of the target material vaporises and ionises. The rise-time of the signal largely depends on the impact velocity of the particle, and the total charge released after impact depends on the impact velocity as well as on particle mass. The pointing direction of the instrument confines the directionality of the particle. The composition of the material is measured through the flight times of the electrons and/or ions after impact. Multiple channels are triggered upon a dust impact, which helps to determine whether an impact is a noise event or not.

**PVDF and piezo detectors** are the still regularly used single-incidence detectors. Their advantages are lower spacecraft resources and larger detector surfaces, but noise characterisation can be more of a challenge. These instruments are generally sensitive to larger particle impacts ($10^{-15} - 10^{-12}$ kg) and basically measure the impact momentum. Successful examples of PVDF-type of detectors are the New Horizons student dust counter, the Cassini High Rate Detector and the Stardust Dust Flux Monitor Instrument (DFMI). Examples of piezo detectors are Rosetta GIADA, and the Bepi Colombo Mercury Dust Monitor (MDM). **Pressurised-cell** type of detectors (also called 'beer-can' type of detectors) like on-board Pioneer 10 and 11, and the early microphones, had specific problems, and are generally not further used. An in depth overview of instruments and dust science is given in Grün et al., 2019.

The last category of instruments can be called **'serendipity instruments'**: after the Voyager 2 spacecraft had detected impacts of dust on the spacecraft body using its **plasma wave instrument** during its Saturn encounter, many more missions have followed where plasma wave instrument data were analysed for dust impacts, e.g. WIND, Cassini, STEREO and most recently, Parker Solar Probe. Particle impacts are measured through the (transient) potential difference that occurs between spacecraft surface charge and antenna, via the impact plasma cloud that is created after particle impact on the spacecraft surface (or on the antenna). This can be detected as a spike in the plasma wave detector data (e.g. Malaspina et al., 2014). A second type of `serendipity measurements' are by high-precision **interferometry missions**, like LISA Pathfinder, where the positions of two free floating blocks are monitored to the nanometer level with an interferometric system (Thorpe et al. 2016). This method is promising for future interferometric missions like LISA.

**Goals and challenges**

Maps of the sky in the infrared, microwave or optical wavelengths exist, as well as sky maps of Energetic Neutral Atoms, but a comprehensive map of our "sky in dust" has not yet been established. Such a mapping, including dust composition, and the link between the source bodies and the measured particles over a wide range of dust sizes - would reveal information about the history of the solar system, processes within, and in analogy, help us understand exoplanet systems and debris disks. Required measurements are composition, impact velocity, and impact direction. One challenge herein is to distinguish the different populations, e.g. interplanetary from interstellar dust, and source body identification from the dust trajectories. Another challenge is measuring the very large (but not very abundant) dust particles, their directions of motion and their compositions, which requires large detector surfaces. Recently, a JUNO mission star camera has observed spallation products of large impacts onto solar panels during interplanetary cruise phase (Benn et al, 2017, Jorgensen et al., 2020). The results are under debate, and the method to be counted as a serendipity observation. Also a challenge, but bringing us in uncharted territory, would be to expand the detection size range towards a `*picodust detector*'. Current detection limits are around 5-10 nm, depending on impact speed. Also, classical spacecraft related constraints exist, like mass, power consumption, data rate, radiation hardness, pointing restrictions (not towards the Sun) and noise and contamination characterization.

**Future instruments tailored for these goals and challenges**

Recent instrument developments are paving the way towards such detections. Not only has the sensitivity of the instruments increased to allow nanometer-sized dust detections, also contamination issues are better controlled, and mass resolutions have increased from M/ΔM = 30 fo Cassini CDA, to M/ΔM = 200, thanks to reflectron-type of mass spectrometers. One of the most promising developments is the "**dust telescope**": a chemical analyser combined with trajectory sensor (accuracy 1°). Similarly, an "**active dust collector**" is a sample return dust collector combined with trajectory sensor, facilitating finding back the impacted particle, its direction and time of impact. Such a trajectory sensor is basically a set of several wire planes, where an electric charge is induced on the wires when a charged dust particle flies through the grids, allowing to determine its direction. New ground-based instruments like the Vera Rubin observatory and the EISCAT-3D radar will improve detections of parent bodies and meteors. In situ dust detectors are currently being built for flying to the Jovian environment on Europa Clipper (Surface Dust Mass Analyser (SUDA) will measure the Europa surface composition), with launch date in 2024. The Interstellar Mapping and Acceleration Probe (IMAP), also scheduled for 2024 has the Interstellar Dust Experiment (IDEX) dust analyser on board, to measure the composition of interplanetary and interstellar dust. Destiny+ will be launched in 2024 with the Destiny+ Dust Analyser (DDA) on board and will visit the active asteroid Phaethon while also monitoring the ISD particles. While SUDA is rather designed for the analysis of low-velocity impacts, DDA is rather tuned towards more sensitivity, with slightly lower mass resolution of 150 instead of 200. Last but not least, the ESA F-class mission "Comet Interceptor" is in preparation for a launch in 2028 to a yet-still-undiscovered (interstellar) comet.

Summarising, the science of cosmic dust in the solar system reveals much more than the dust alone. It comprises compelling subsurface and surface science, processes in space, physics of space plasmas and magnetic fields, history of the solar system, etc. and space instruments provide complementary information to meteors, meteorites and astronomical observations. A legacy of more than half a century of cosmic dust instrumentation has lead to increasingly powerful instrumentation. A future Dust Observatory would open a new window in Dust Astronomy.

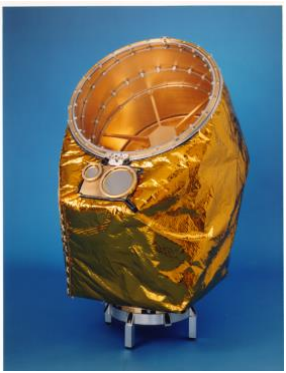

*Fig 5.5. Example of an instrument for dust investigations*

### 2.3.2. Mass spectrometry

The future challenges in mass spectrometry for the next fifty years are the demand for increased analytical performance and sampling systems that can adapt to challenging extreme environments.

Increased mass resolution is a prerequisite for future mass spectrometry applications, improved separation techniques such as electrophoresis and comprehensive two-dimensional gas chromatography, and most notably position specific isotope analysis are required to meet the analytical rigor needed to explore the origin and evolution of the Solar System and the life within. Future missions that will require the highest precision in mass spectrometry include exploration of the surface of Venus, probes into the atmospheres of the ice giants, and landers and ocean probes to the Ocean Worlds of the outer Solar system.

**Enabling technologies: current status and future capacities needed**

Each of the challenging missions listed above brings with it a unique requirement for sampling the target environment: i) Venus' extreme surface heat and sulfuric acid clouds make sampling of important volatiles such as noble gases a challenge, ii) sampling of the atmospheres of the ice giants requires sampling systems that must deal with a broad range of pressures and provide a uniform vertical sampling of gases and aerosols, and iii) sophisticated organic analysis on icy moon surfaces or an interior ocean present both analytical and sampling challenges.

Rosetta ROSINA (Balsiger et al., 2007) has clearly demonstrated the importance of improved mass resolution to elucidate the hidden clues in complex organic mixtures. Future missions to Mars (Brinkerhoff et al., 2013; Arevalo et al., 2015; Goetz et al., 2017; Li et al., 2017; Goesmann et al., 2017) and to Galilean satellites (Wurz et al., 2012; Brockwell et al., 2016) strive to provide the needed mass resolution to sort out the CHON content of organics with masses up to mass 200 u offering much improved sensitivity by using time-of-flight mass spectrometry techniques.

Efforts are under way in France and the USA to push the envelope of available mass resolution for future missions (Briois et al., 2016; Selliez et al., 2019; Arevalo et al, 2020). Improved separation techniques are being studied with NASA R&D funding for lander and flyby missions to Ocean Worlds: i) Two-dimensional gas chromatography (Blase et al., 2020; Glein et al., 2019) and ii) microcapillary electrophoresis (Mathies et al., 2017).

Two technology thrusts are anticipated in the coming decade:

i) Sampling systems that can sample in extreme remote environments such as the atmosphere (and surface) of Venus (Gruchola et al., 2019), the depths of the ocean of Enceladus, and the atmospheres of the ice giants (Vorburger et al., 2020).

ii) New technologies to support the bold new analytic methods required to search for life from the Oceans of Europa and Enceladus (Hand et al., 2017; Cable et al., 2017; and Engeinbrode et al., 2018). Such methodologies are set forth in An Astrobiology Strategy for the Search for Life in the Universe. National Academies of Sciences Engineering and Medicine. (2019). Washington, D.C.: National Academies Press. https://doi.org/10.17226/25252 and in Neveu et al., 2017.

Paradigm changing opportunities are already being explored and discussed, especially in the search for life elsewhere in the Solar system.

i) Position sensitive isotopic analysis (psia) has been demonstrated in the laboratory to give important information about the origin of simple molecular compounds (Eiler et al., 2017) and efforts are underway to adapt this to spaceflight.

ii) Nanopore sequencing, although not strictly a mass spectrometry technique has been tested onboard the International Space Station (Castro-Wallace et al., 2017) and is being further advanced for planetary applications (Bywaters et al., 2017). It seeks to exploit the reliance on a polymer with repeating charge (polyelectrolyte) as a means to store and pass on genetic information could be a universal feature of life (Benner, 2017).

**Main conclusions and suggestions for future developments**

Table 5.2 below summarizes new analytical techniques discussed herein.

| Technology | References | TRL |
|---|---|---|
| High resolution and sensitivity mass spectrometry | Wurz et al. (2012); Brinkerhoff et al. (2013); Brockwell et al. (2016) | 6+ |
| Ultra high resolution mass spectrometry | Briois et al. (2016); Arevalo et al. (2020) | 3 |
| Comprehensive two-dimensional gas chromatography | Glein et al. (2019) | 2 |
| Microcapillary electrophoresis | Mathies et al. (2017) | 5+ |
| Ice giant descent probe sampling | Vorburger et al. (2020) | 2 |
| Position sensitive isotopic analysis | Eiler et al. (2017) | 2 |
| Nanopore sequencing | Bywaters et al. (2017) | 3 |

*Table 5.2. Some of the promising new analytical techniques for mass spectrometry.*

## 2.4. Life detection devices

In recent decades, planetary exploration has increasingly focused on the in-situ astrobiological research, in addition to the classic geological and environmental approaches on the surface. So, the current missions and programs of the largest space agencies include the detection of traces of life among their direct objectives.

With the exception of the Vikings, the most recent missions have faced this challenge from a cautious approach, just determining or characterizing the present and past habitability conditions on the planet (which, of course, is not indicative or proof that the planet was -or is- inhabited); or making use of spectroscopic, microscopic or chromatographic techniques, essentially, to identify signatures of organics or of more ancient microbial life[1].

However, given the nature of the objective, and the importance of the discovery of potential life outside Earth, the space agencies are aware of the difficulty and degree of certainty that must be achieved to make such an uncontroversial claim public, with the corresponding level of the technological maturity that the missions would require (for that same reason, NASA Perseverance does not include the search for living microorganisms in the present Mars among its scientific objectives).

---

[1] As it is the case of the Rosetta-Philae's *Cometary Sampling and Composition* (COSAC), ExoMars's *Mars Organics Detector/Mars Organic Analyzer* (MOMA), or the *Scanning Habitable Environments with Raman and Luminescence for Organics and Chemicals* (SHERLOC) onboard NASA's Perseverance rover, for example, instruments limited to the ambiguous detection of amino acids and other related organics, which may not be conclusively indicative of a biogenic origin.

Nevertheless, the different agencies are becoming more dynamic in this endeavor, bringing together the diverse disciplines and experience, defining approach strategies, and fostering and maturing the technologies necessary to face this important and far from easy challenge: the search for evidence of life beyond our Earth.

For the moment, in the programs that are currently underway, the largest agencies are opting to select the best candidates, the most suitable Martian rock or soil samples to harbor evidences for life (in the specific case of Mars), and bring them to Earth for confirmation without doubts. However, this strategy could not be, in the medium-long term, the most efficient. So, in the recent years, numerous research groups are developing and maturing different technologies, combining some of them, as well as defining strategies that leave no room for doubt in identifying life in-situ, as happened in the past.

However, the big question remains open: what level of chemical complexity could be considered definitive evidence of the existence of life (past or present)?

To try to answer this question unambiguously, as mentioned, an adequate approach to the problem is necessary, defining a minimum set of criteria and observations that allow the identification of universal life forms, and developing a new generation of instrumentation that may provide the relevant data.

Some of the most promising research lines propose to perform potential indigenous life growth in controlled environments, bioaffinity-based systems for finding well-preserved molecular structures, extraction of organic molecules in liquid suspensions, or the detection of non-volatile complex compounds (low molecular weight biomolecules, biopolymers, as well as macromolecules, supra-molecular complexes and quasi-cellular morphologies), as approaches to identify different levels of prebiotic/biotic chemical and structural complexities from a non-Earth-centric perspective.

Additionally, this instrumentation must be able to survive the most extreme planetary environments, with a reduced size and weight, as automated as possible, and with moderate power consumption. Without a doubt, all of this sets out significant technological challenges.

Various technologies can meet these demanding limitations and capabilities. One of these promising ones is the use of nanopore-based devices for the detection of organic compounds of biogenic origin, which are currently used to detect and sequence nucleic acids, some derivatives composed of non-standard-bases, as well as potentially xeno nucleic acids. All this, using very compact and robust devices that can fit the strict requirements of the planetary exploration missions.

Another promising approach is the combination of various techniques, as previously discussed, to unequivocally resolve the existence of these potential life forms. As an interesting example of complementary techniques is: microscopy, to identify cell-like ultrastructures and morphologies, which can help resolve the evidence for life at the microscale; Raman spectroscopy, for detecting universal intramolecular complexity, in particular, revealing the non-covalent bonds and atomic composition, resolving the 3D secondary and tertiary polymeric structures, and identify other signatures of organic molecules of different sizes and complexities, not restricted to the know biochemistry operating in our planet; and a bioaffinity-based system where biological molecules are used to capture other ones, being able to identify the nature and structure of those complex targets detected (or part of these) by means of the lock-and-key principle.

Individually, these techniques already have a high level of technological maturity and flight qualification, so the synergistic combination of them will reduce risks and potential identification gaps, as well as will provide an excellent method of addressing the problem presented, allowing to cover a wide spectrum of targets in the search for (bio)chemical complexity in the Solar System, as well as its (dis)similarity with the biochemistry of terrestrial life.

This novel instrumentation and techniques, or a combination of them, will undoubtedly add important value to the future planetary missions, unequivocally identifying those universal forms of life outside our planet.

# 3. Mission architectures for the future

## 3.1. Pathways and distances to the different known Solar System destinations

Delivering instrument platforms to the different solar system destinations to perform scientific operations involves a chain of space vehicles that first reach an Earth-Moon orbit, then fly to interplanetary space and execute planetary or small bodies fly-bys, before finally going into orbit around their planet, moon or small body destination, and in some cases landing on them. Figure 5.6 illustrates the diversity of paths followed to this day by spacecraft that have flown to well-identified solar system destinations. The differences in numbers of visits between "near-by" destinations (Moon, Mars and Venus), more distant planets (Mercury, Jupiter and Saturn) and solar orbit, and those located in the outskirts of the solar system (ice giants, Kuiper belt objects and heliospheric boundaries) is striking!

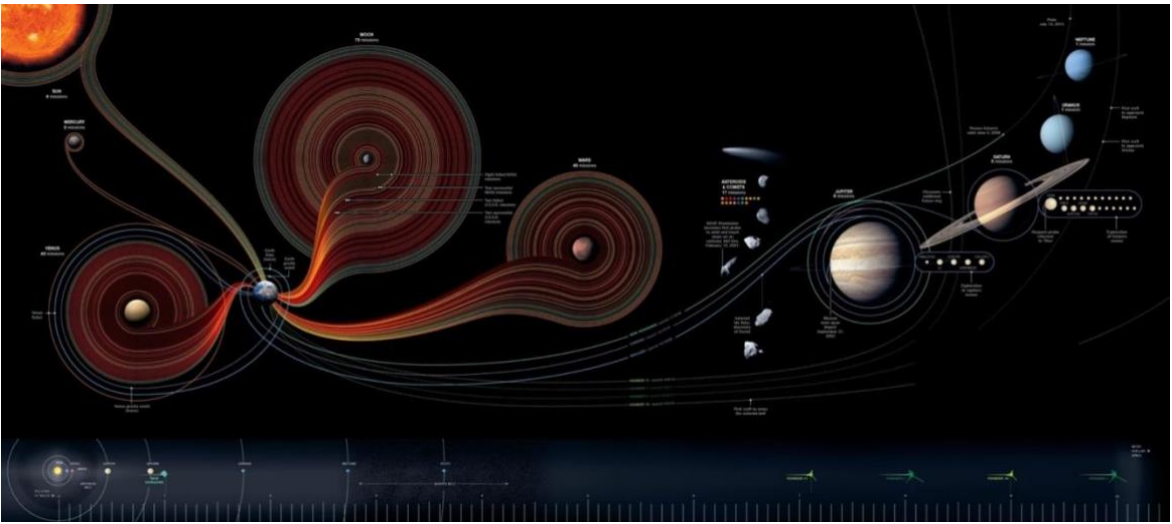

*Figure 5.6: Overlay of the diversity of paths followed by our spacecraft to solar system destinations until this day. The difference in number of flights between nearby and distant destinations is very large, showing that our exploration of the diversity of solar system objects is still heavily biased by their distance to us (find reference).*

From an operational viewpoint, distances along these long and exciting journeys are counted less in millions of kilometers or Astronomical Units, than in travel times and Delta V budgets: these are the parameters which determine the duration of planetary missions (with important consequences on their costs) and the fraction of mass to be allocated to propellants rather than to the spacecraft and its different platforms, which determines in the end the mass that can be allocated to scientific instruments for a given resource envelope. Figure 5.7 is a schematic graph giving approximate Delta V "distances" to a host of solar system destinations. The total "distance" to each destination is shown next to it in green, while the "length" of each segment is indicated in blue along the different paths.

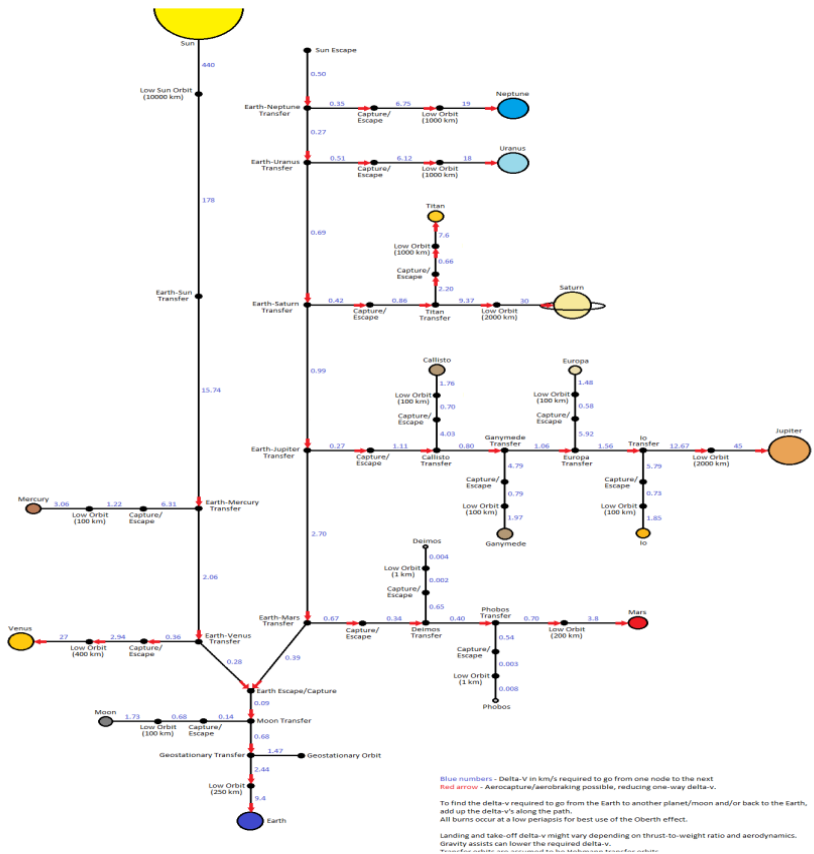

*Figure 5.7: Schematic map of distances to the different destinations to the solar system, in units of Delta V. Blue numbers show the approximate cost of each segment, and green numbers near each destination the total distance of each target to Earth. (this figure should be edited or re-drawn, and green numbers added).*

Variation of these "distances" across the solar system are huge. Gravity assist helps reduce these Delta V distances, but at the cost of a longer travel time: in the end, mission designers have to choose between these two "distances", travel time and Delta V, and find their optimal location in this 2-D parametric space given the resources allocated.

As can be seen from the Delta-v map of the Solar System reproduced above, our ambitions for the exploration of this Solar System are constrained by the available Delta-v of our rockets. Delta-v represents the total change of speed needed over a mission – it is a measure of the impulse per unit mass required for a journey. For a rocket, it may be calculated using the Tsiolkovsky ideal rocket equation. Detailed examination of the map shows that we have been limited in our explorations by current technological limitations on Delta-v capability. The map does not show the important gains to be made by gravity assists but nevertheless shows clearly the relevant limitations. Note that a sample return mission requires us to travel in both directions along a specific route. All the targets listed are essentially in the plane of the ecliptic. Future missions will therefore require significant advances in this area.

It is immediately obvious that the mass to be delivered must be minimised. New technologies such as ion drives which maximise the velocity of the ejecta (although the continuous burn makes the flight dynamics more complicated) are obviously worth investigating, and have been applied particularly to Mercury (ref) where the increasing Solar flux makes the combination of Solar Electric propulsion highly desirable. Non-rocket approaches such as tethers, Solar sails, plasma sails, space elevators, lasers and launch guns (refs needed) could help minimise the amount of fuel that is needed during the journey, and hence reduce the launch mass, with the limitation that the low thrust provided by most of them and the need to compensate for it by long-duration propulsion phases take these propulsion solutions far from the optimal transfer case and may generate significant additional DeltaV costs. It is quite possible that by 2061 new approaches will be developed, and that could represent a significantly disruptive improvement. However, it is probably unlikely that a new technology will reach practical large-scale maturity in the period under consideration. The proposed Interstellar probe (McNutt et al in this book) represents the limit of currently available technologies.

## 3.2. Articulation between interplanetary carriers and science platforms

As time progresses, planetary missions performing initial surveys of the different systems and objects, using mainly interplanetary and planetary orbits and performing remote sensing studies of objects and in situ observations of the space environment, will make way for more specialized missions targeting specific objects and addressing focused science objectives. This major trend is already dominant for Moon, Mars and Venus missions, and emerges in the next wave of missions to outer planets, moons and small bodies. It has deep consequences on the architecture of planetary missions: while they must always "pay" for the travel time and Delta V costs of their chosen destination, via the use of a launcher and an interplanetary carrier, their science operations are performed more and more by specialized science platforms operating the appropriate instrument suites at their chosen destination. These platforms and their instruments are designed to operate in the specific environment conditions of these destinations. As distances to Earth increase, these environments offer more and more "extreme" conditions, and require "adaptation" to these environments in a near-Darwinian sense: the adaptive pressure of this evolution on the design of science platforms and instruments is driven by the selection processes leading to the choice of instruments, and to the ever-increasing demands on the quality and scientific value of the scientific data to be returned to scientists for analysis.

It is easy to understand in which direction this selection pressure drives the architecture of planetary mission, since this evolution can already be observed in the missions currently in operation or in preparation: missions operating on Mars surface use rovers with electric propulsion powered by radio-isotope devices to navigate martian landscapes, while NASA prepares for the DragonFly mission to deliver a drone to explore Titan's surface and atmosphere. In line with this evolution, the interplanetary spacecraft that deliver the different platforms at their operational locations will naturally specialize more and more on their "core" mission of carrying and delivering a payload to a destination. Along the paths of each mission illustrated in figures 5.6 and 5.7, this "transportation" function is complemented by vehicles dedicated to specific critical segments: launchers between Earth's surface and orbit, Entry, Descent, Landing and Ascent systems (EDLA) systems between interplanetary or planetary/moon/small body orbits and the surface or atmosphere of a target body.

As a natural consequence, segments of the mission performing solely or mainly transportation and navigation functions with only loose connection to science operations could be accomplished by operators offering solely transportation services, be they space agencies or commercial operators: this other heavy trend is also already observed in the current developments of Moon operations, and witnessed by the interest of some commercial companies for Mars exploration. This emerging trend may be beneficial to the science-driven dimensions of planetary operation addressed in this book if it has the effect of increasing the number and frequency of flight opportunities for scientific operators and their platforms. But this benefit would be real only if access to space and to the different destinations of the solar system remains affordable.

### 3.3. Missions and functions of science platforms.

Once at destination, science operations are performed by general or specialized science platforms, whose functions and operation constraints are guided by the environment in which they have to operate:

- Orbiting platforms can generally combine two types of scientific missions: remote sensing of atmospheres, magnetospheres, surfaces and interiors of planetary bodies, and in situ characterization of the space environment; until now most of these platforms remained attached to their interplanetary carrier, and benefited from its functions and resources (power, orientation, navigation, telecommunications, protection against the space environment); in the future, they may evolve towards independent stand-alone platforms delivered to their final orbit by an interplanetary carrier;
- Atmospheric science platforms are used to characterize planetary and moons atmospheres, their composition, dynamics and nebulosity. Their delivery may involve entry probes that take care of their hypersonic entry, deceleration, stabilization and deployment;
- Surface science platforms can also perform atmosphere characterization near the surface of a planetary body, but mostly perform science operations in geology, geophysics, geochemistry, and astrobiology when relevant. Their delivery to a surface requires an Entry, Descent and Landing System (EDLS);
- Finally, sample return operations require a platform that can explore the sampling site, characterize it, select samples, and collect them. Their delivery to a surface and the return to Earth of the sample requires an Entry, Descent, Landing and Ascent system (EDLA).

### 3.4. The role of small and multiple platforms

The evolution of the science objectives of these platforms, combined with mass allocation constraints, drive their evolution with time in two ideally complementary directions: miniaturization, with the development of small, low-cost platforms less demanding in terms of on-board resources, and observation systems using several platforms.

While the prospects for an increasing capacity of small platforms are described in section 5, let us review here the scientific drivers for multi-platform measurements. Generally speaking, evolution towards multi-point measurements of planetary bodies and environments is driven by the need for measurements to cover their spatial and temporal variability, which can only increase with progress in our knowledge mission after mission. This need for multiple measurements drives the design of multiple-platform observing systems in nearly all areas of planetary observations:

- For magnetospheric and space environment research, multi-point measurements have become increasingly important to capture the extreme variability of these media, to disentangle spatial from temporal variations of quantities observed along the trajectory of the spacecraft, and to explore the extreme diversity of spatial scales involved in the dynamics of space plasmas. This evolution towards multipoint measurements and multi-platform missions has dominated Earth magnetosphere missions for two decades already with CLUSTER, MMS and other missions. Given the comparable (and sometimes larger) complexity of other magnetospheres, such as Jupiter's for instance, flying multi-platform missions to the most complex of these magnetospheres should be an important trend for the coming decades;

- For atmospheric sciences, using several entry probes will be useful to compensate for the regional variations of chemistry, weather and dynamics, as well illustrated by the scientific analysis of the Galileo probe; another scientific use of multiple platforms is the deployment of two-way communication links, by means of which atmospheric and ionospheric profiles can be obtained by radio occultation with a broad diversity of geometries (section 2.2.3 above).

- Finally, for geosciences and astrobiology, some science objectives require the detailed exploration and characterization of a single site or limited area, while others call for observing the regional and global variability of planetary surfaces and sub-surfaces. Observing the diversity of phenomena at the regional scale may be performed by mobile platforms such as rovers, balloons, aircraft or drones. The global characterization of some geophysical phenomena, such the internal structure and dynamics of planetary bodies, requires in some cases the deployment of a global geophysical network to perform seismic, magnetic and other geophysical network observations.

For planetary applications, constraints on the mutual positioning of the different platforms of a multi-platform system are generally weak: mutual positions need only to be known, not controlled, with an accuracy significantly better than the lowest spatial scales of variability.

## 3.5. Architecture of missions "on alert"

Besides the large number of identified destinations, the solar system also offers "unknown" destinations to explore: these objects which have not yet been detected or identified, and for which the "classical" approach of mission planning and the corresponding routes do not work. These destinations are of the utmost scientific and programmatic interest (see Dehant et al., 2020) as key elements to build a more comprehensive inventory of solar system objects, and particularly to understand the origins of the solar system and its connections with its galactic environment. Two examples of first order importance are interstellar objects and dynamically new comets.

The first identified interstellar object, 2017 1I/'Oumuomuo, was detected in 2017 as a cigar-shape small body, the size of a skyscraper, crossing the solar system on a hyperbolic trajectory (Fitzsimmons et al., 2017; ISSI Team report, Nature Astronomy, 2019): see Figure 5.8. This kind of object is of the utmost interest, as it has likely formed in a different protoplanetary disk and scattered to interstellar space by gravitational interactions with planets and small bodies in that disk. (Pfalzner and Bannister, 2019) studied this class of objects and suggested that, given their transport between different planetary systems and protostellar disks in the Galaxy, they might even play a role as seeds in the formation of planetesimals, these "building blocks" out of which planets finally accrete. Though the probability of occurrence of such an object in the Solar System is low, it is not negligible and worthy of consideration: a recent study by (Seligman & Laughlin 2018) shows that, by using the detection power of the Large Synoptic Survey Telescope (LSST) one should be able to detect one such object about once every ten years: enough to try and plan a mission to visit them!

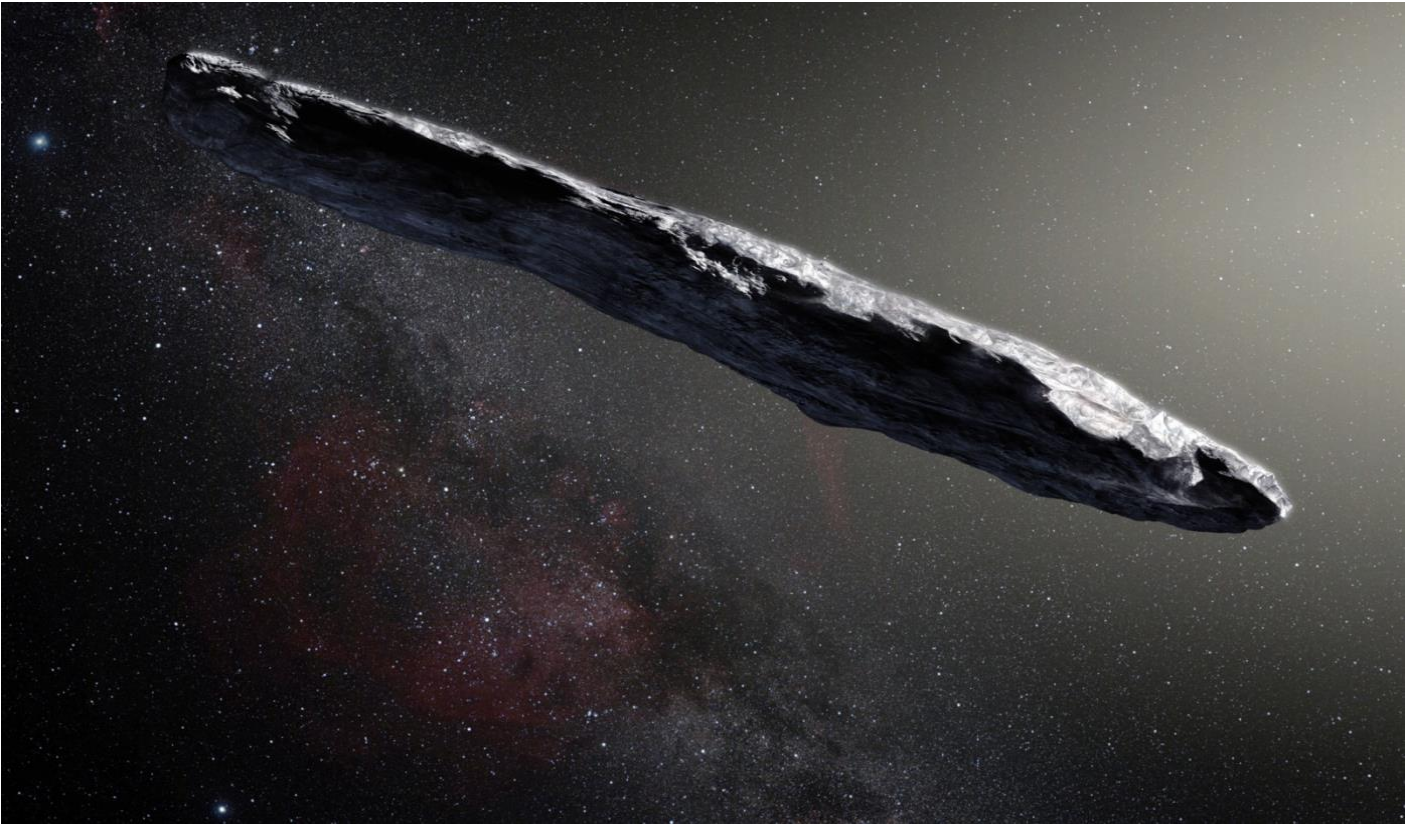

*Figure 5.8: a representation of the first ever detected interstellar object, 2017 1I/'Ouamuomuo. This class of object, likely formed in the protoplanetary disks of other planetary systems, is of the utmost scientific interest, but their occurrence frequency in the solar system, even with the most powerful telescopes like the LSST, is only on the order of one every ten years. Courtesy ESO/M. Kornmesser.*

Dynamically-new comets are another class of objects whose location and trajectory are unknown, as they are "new" in the sense that we observe them during their first visit to the inner solar system. Free of degassing activity and of weathering by the solar system space environment, they offer a fully pristine material, particularly with its fresh surface ice cover, which is as representative as possible of the conditions in which they formed during the assembly of the solar system. Their detection in view of visiting them with a spacecraft must be done as early as possible along their inbound trajectory in the outer solar system, in order to be able to plan a mission encountering them not too far from the Earth's orbit.

These categories of newly discovered objects require a special mission architecture and planning. Given their scientific interest, these missions "on alert" should progressively become part of the overall planning and coordination of space agencies, to take the best possible advantage of the few opportunities offered by new detections. The recently selected Comet Interceptor F-class mission of ESA's science program, currently under implementation, should be a template for more missions of this kind in the 2061 time frame. Figure 5.9 illustrates its mission profile.

As the figure shows, this very innovative mission scenario combines the challenges of a mission "on-alert" and of a multi-point investigation of the cometary environment: while spacecraft A (the main spacecraft, delivered by ESA) passes at a safe distance of about 1000 km upstream of the comet in the solar wind, and serves as the data relay satellite for the other two platforms, spacecraft B1 (delivered by JAXA) visits the inner coma with a high survival probability and spacecraft B2 (delivered by ESA) is targeted for an encounter with the nucleus, with a high risk of being destroyed after having sent back close-in images and other data on the coma close to the nucleus.

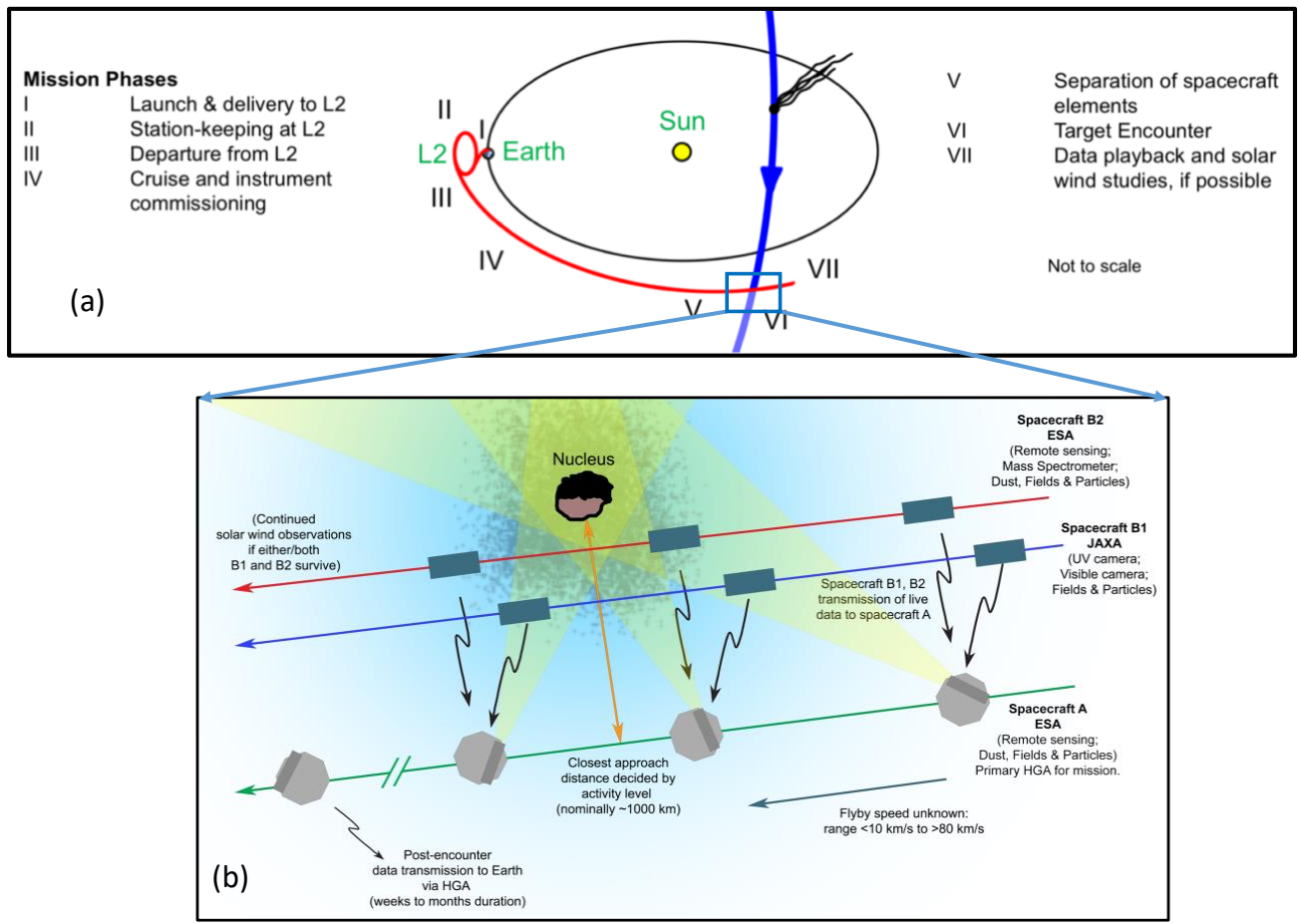

*Figure 5.9: Summary mission profile of ESA's Comet Interceptor mission to a dynamically-new comet. (a) Mission plan to meet the comet: first, parking on an L2 Lagrangian point halo orbit, waiting for detection of its target and injection on a near-Earth encounter trajectory heliocentric orbit; (b) scenario of the three-platform encounter with the comet.*

This mission offers a perfect example of complex architectures, combining in that case the advantages and challenges of a mission "on alert" and of a multi-platform observation system.

Some events, such as the arrival of long period comets or Interstellar asteroids (Refs) are intrinsically non-predictable, but nevertheless can provide scientifically compelling possibilities for studies (see discussion in Chapter 4). Similar Delta-v considerations apply to these targets of opportunity as the previous chapter. However, an extra problem is raised by the fact that these targets will in general be significantly out of the plane of the ecliptic, meaning that the budget is both large and unpredictable. However, the payloads required may not themselves be very large or demanding since the chances of prolonged rendezvous or landing are at this stage very slim. This is however a clear example of a mission where small expendable smallsats may enhance the mission science profoundly.

## 4. System-level technologies to fly there… and return

### 4.1. Introduction

Some of the future deep-space exploration missions will face more extreme deep-space environments. The diversity of exploration targets and the prolongation of exploration time will certainly put forward more stringent requirements for deep-space spacecraft.

For spacecraft system design and analysis, as the overall complexity of the probe system becomes higher, including orbiters, landers, ascenders, submersibles, rovers and extraterrestrial surface vehicles, the interface between the systems becomes more complex and interoperability becomes more important. This inevitably requires the use of, complex system modelling and system simulation with the help of artificial intelligence (AI) methods to design flight schemes and spacecraft system schemes.

For in-space propulsion, *there is no single propulsion technology that benefit all mission types. The requirements vary widely due to their intended application*. In addition to continuing optimisation of the specific impulse and other performances of the chemical propulsion, *development of non-chemical propulsion system such as low-power and higher-power electric propulsion, and nuclear thermal propulsion, will have the broadest overall impact on enabling or enhancing different types of missions. Development of cryogenic chemical propulsion* is also of potential interest when large masses have to be moved to and from planetary surface such as for manned missions to Mars. For some specific fly by deep space missions with small probes, special propulsion methods such as solar sails and electronic sails can also be developed.

For advanced power, solar energy is widely used for spacecraft systems or even surface stations and rovers when the received solar energy density is high enough, in average. Nuclear energy must generally be used in other cases such as deep space flights. Missions to explore the surface of an extra-terrestrial body (such as Moon and Mars) could also consider on a case by case basis more specific ways of generating electricity in solar power stations based on surface temperature difference on the moon, or wind energy on Mars. And maybe one could use the energy of underground ocean flows beneath the ice to generate electricity for submersibles.

For telecommunications, improvements can be made either on the spacecraft or on the ground systems side, with the same goal of increasing the received power, and therefore the data rate, from the spacecraft.

Autonomous navigation technology should be vigorously developed: since deep space missions can only rely on astronomical navigation, the autonomy of the spacecraft, as well as autonomy control and health management, are very important.

At last, in order to improve the intelligence of deep space probe, it is necessary to vigorously develop advanced computing capabilities, and IA applications to science missions in areas such as such as big data, cloud computing, expert knowledge system and other technologies to improve the autonomous online processing ability of the probe.

## 4.2. Spacecraft system design and analysis

### 4.2.1. Introduction

According to the mission's scientific objective requirements, spacecraft system design and analysis is mainly to plan the corresponding spaceflight mode, determine the mission architecture and mission operation scenario, complete the system function and scheme design of each spacecraft. Requirements of planetary missions on spacecraft system technologies

As exploration destinations expand and increase, the types of exploration missions become more complex and diverse. Mission architectures for the future can be divided into two kinds: human mission and robotic mission. Human missions, which are complex and expensive and require long development times, will be limited to landing on the moon or Mars even by 2061.

Figure 5.10 illustrates the primary transportation Design Reference Missions (DRMs) results of Human Space Flight Architecture Team (HAT) coming from NASA/Johnson Space Center. It reflects selecting different destinations used to drive transportation systems requirements and assess impacts of changes in mission assumptions. Figure 5.11 shows the evolution of key assumptions that drive transportation system performance. In order to make different types of platforms from the transportation system of the future will be cooperate working, the optimization design of their performances and functions will be very important. It also can be seen from Figure 5.11 that different flight modes lead to different design allowances of various spacecraft and directly affect the selection of launch vehicle, and the cost and risk of the project.

## Select destinations used to drive transportation systems requirements and assess impacts of changes in mission assumptions

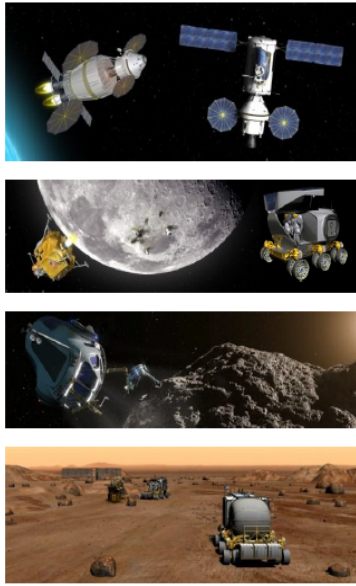

| Proposed Status | ISECG | DRM ID | DRM Title | Dest. |
|---|---|---|---|---|
| Cycle-C | N | LEO_UTL_2A | LEO Utilization - Non-ISS | LEO |
| Cycle-C | Y | CIS_LP1_1A | Lunar Vicinity - EM L-1 | E-M L1 |
| Cycle-C | Y | CIS_LP1_1B | Lunar Vicinity - EM L-1 DSH Delivery | E-M L1 |
| Cycle-C | Y | CIS_LP1_1C | Lunar Vicinity - EM L-1 with Pre-deployed DSH | E-M L1 |
| Cycle-C | Y | CIS_LLO_1A | Low Lunar Orbit | LLO |
| Cycle-C | Y | LUN_SOR_1A | Lunar Surface Polar Access - LOR/LOR | Moon |
| Cycle-C | Y | LUN_CRG_1A | Lunar Surface Cargo Mission | Moon |
| Cycle-C | N | NEA_MIN_1A | Minimum Capability, Low Energy NEA | NEA |
| Cycle-C | Y | NEA_MIN_1B | Minimum Capability, Low Energy NEA with Pre-deployed DSH | NEA |
| Cycle-C | N | NEA_MIN_2A | Minimum Capability, High Energy NEA | NEA |
| Cycle-C | N | NEA_FUL_1A | Full Capability, High Energy NEA with SEP | NEA |
| Cycle-C | Y | NEA_FUL_1B | Full Capability, High Energy NEA with SEP and pre-deployed DSH | NEA |
| Forward Work | N | MAR_PHD_1A | Martian Moon: Phobos/Deimos | Mars Moon |
| Forward Work | N | MAR_SFC_1A | Mars Landing | Mars Surface |

*Figure 5.10. The Primary Transportation DRMs Results*

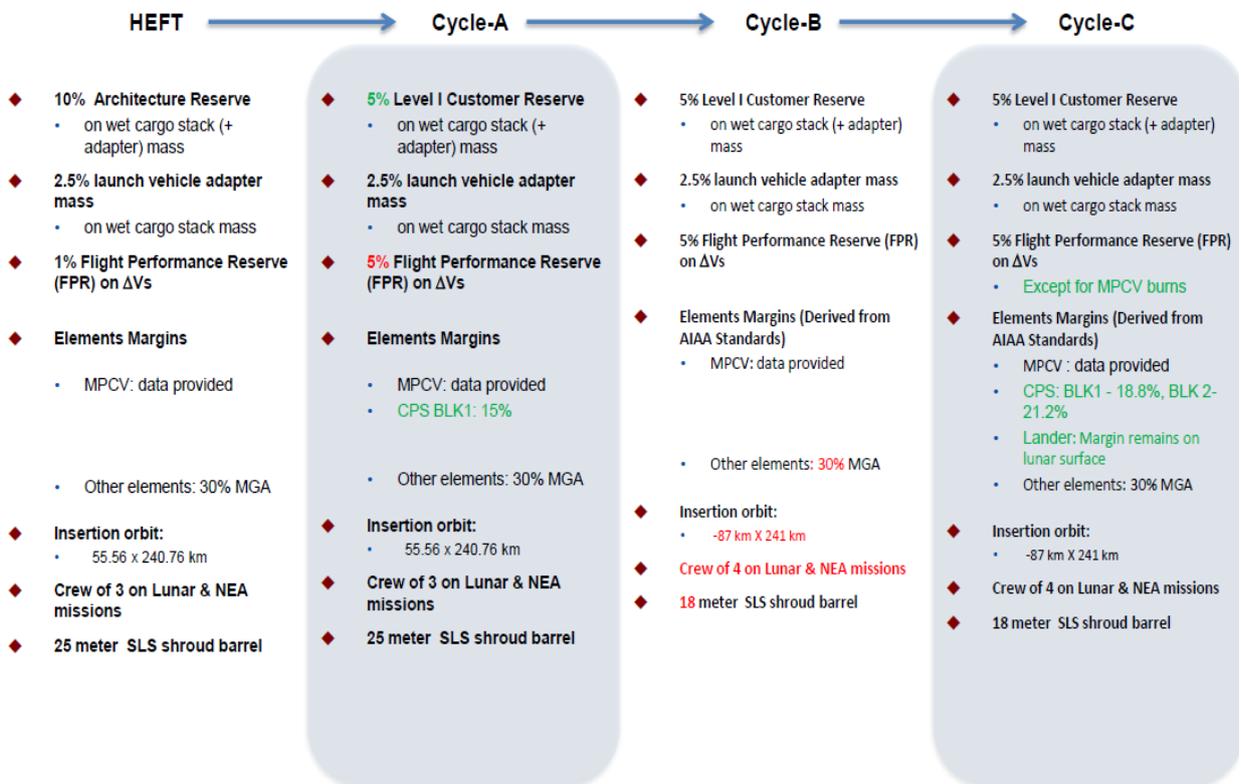

*Figure 5.11. Evolution of Key Assumptions that Drive Transportation System Performance*

## 4.2.3 Enabling technologies: current status and future capacities needed

Traditional system engineering (TSE) design is based on a format of documentation which is responsible for the exchange of information and the management of interface requirements between system, subsystem, device and instrument. However, TSE design is obviously unable to meet the requirements of future missions. With the increase of the types, scales and complexities of spacecrafts, interfaces between the different spacecraft become more and more complex, and the difficulty of information management becomes larger. Therefore, document-based management needs to be changed to a system engineering design method based on digital models, called Model-Based Systems Engineering (MBSE).

The MBSE approach can be used to build mission models based on different destinations, produce models based on different spaceflight modes, cost models based on different architecture products, risk models based on different TLR levels, and so on. Figure 5.12 illustrates how the analysis process of the MBSE approach helps the project leader and engineers to formulate an optimized system mission and spacecraft design. The MBSE approach will need to be further developed and evolve to face the more complex space missions that will fly before 2061. For example, it should take advantage of the development of virtual simulation systems, expert knowledge systems, cloud computing and big data technology.

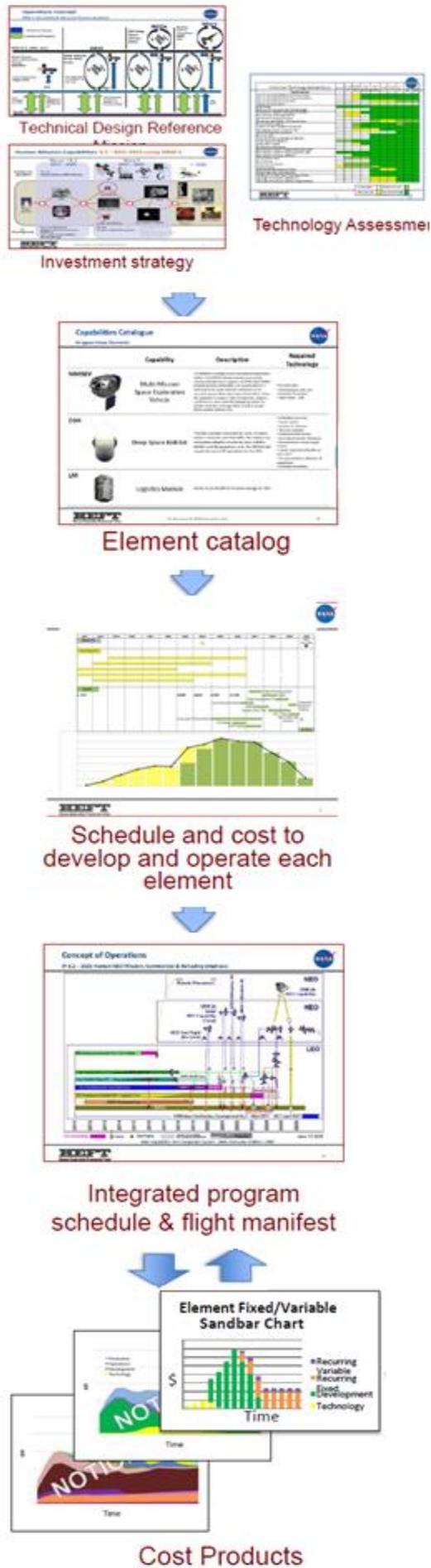

*Figure 5.12. Architecture Products, Technology and cost analysis based on MBSE.*

## 4.2.4. Suggestions concerning future developments

With the development of deep space exploration science and technology, spacecraft design and analysis methods should become more convenient, detailed and easy to use. The development of space industry informatization and digitization precisely provides the possibility to promote the progress of spacecraft design and analysis methods. Considering the wide application of artificial intelligence technology, future spacecraft design will be more and more intelligent and humanized.

## 4.3. Advanced Electric Propulsion

Electric propulsion systems of electro thermal or electrostatic plasma acceleration types have already been used on hundreds of Earth orbiting satellites and on some space probes. The ESA/JAXA BepiColombo mission launched in October 2018 relies on electric propulsion in cruise, employing 4 gridded ion thrusters. This mission is set to arrive at Mercury in late 2025.

In the last years, advances in solar power generation systems have increased the total amount of available on-board power, and electric propulsion-based orbit transfers using 50kW or more of electric power are becoming realistic. Simultaneously, there is a need for low power (1-500W) electric propulsion systems for the exploding and disruptive market of small satellites, where the thrust level is in the order of microNewtons. Several product developments using different types of electrostatic propulsion technologies and of propellants are currently going for such applications in the world. This trend is driven by an extreme cost reduction and the possibility for mass production via for example the use of the standardized CubeSat technology, traded against quality and long lifetimes.

Several types of electric propulsion systems have been or are used in space, notably : electrothermal thrusters (arcjets and resistojets types ) (65%)   electrostatic thrusters (gridded ion engines, Hall thrusters, Electrospray types) (35%) and magnetoplasma rocket (VASIMR still at a development stage). Between these thrusters, some provide higher thrust and some provide higher specific impulse thus they are not directly competitors but rather complementary products.

The near-term challenge for propelling large spacecraft and interplanetary probes is to develop long life span, high-power thrusters (> 10 kW). Hall thrusters of 20kW have been developed recently (Zurbach et al., 2011) and a nested-channel Hall thruster is under development aiming to operate above 100 kW (Hall et al., 2014). In addition to a high-thrust level, dual-mode or multi mode capability (i.e. different values of specific impulse over thrust ratio for the same input power) is also of relevance to optimize mission profiles. Another critical research area for ionic/plasmic propulsion is to identify alternative propellants to xenon to reduce the overall cost and improve the safety of using electric propulsion.

Even if thrusters such as Hall effect thrusters have been extensively studied since their invention in the 1960s, several plasma processes that have direct relevance to the thruster performance and lifetime are still poorly understood. Today, the design and development of Hall effect Thrusters is still semi empirical and requires long and expensive life tests. There is a need to better understand crucial plasma processes occurring in the complex magnetized plasmas at the core of Hall effect thrusters (Adamovich et al., 2017, Boeuf et al. 2017; Kaganovich et al., 2020) such as electron transport, interaction with walls and erosion, and to address the question of alternative propellants. One also needs to better understand plasmas in the real architectures of Hall effect thrusters, to develop 3D numerical tools for the simulation of such problems and to make them available to the industry, and to use these tools to improve the efficiency of existing products. This is what will provide the foundation for breakthroughs in the designs of future electric thrusters. Finally, another critical point is the influence of ground-test facility effects on performance.

## 4.4. Advanced Power

## 4.4.1. Introduction

The science and robotic exploration programmes of many agencies include future missions to the outer planets, and missions with planetary landers and rovers. For these missions, nuclear power sources are a key enabling technology. It is also widely acknowledged that human presence on Mars would require nuclear power, probably reactors. (*Stephenson and Blancquaert, 2008),* although geothermal energy generation has been suggested as a longer term solution *(ref needed).*

It is not the pupose of this article to provide a comprehensive review of Nuclear power in space. It is recommended that the interested reader consult the review by Stephenson and Blancquaert, (2008), on which this article draws heavily. We will however consider where breakthroughs are necessary for missions under consideration in the 2061 timeframe.

## 4.4.2. Conventional Solar Power Generation

While solar energy generation is an excellent and well developed solution for space exploration in the inner solar system, beyond Jupiter it is clear that different tecnology is needed. ESA's *Rosetta* required 64 m2 of solar array to just survive at 5 a.u.

Where atmosphere and/or dust are present, even the daytime power can be unpredictable. In Martian dust storms, optical depths ($\tau$) of greater than 4 have been measured. This equates to one tenth the sun power of Earth orbit.

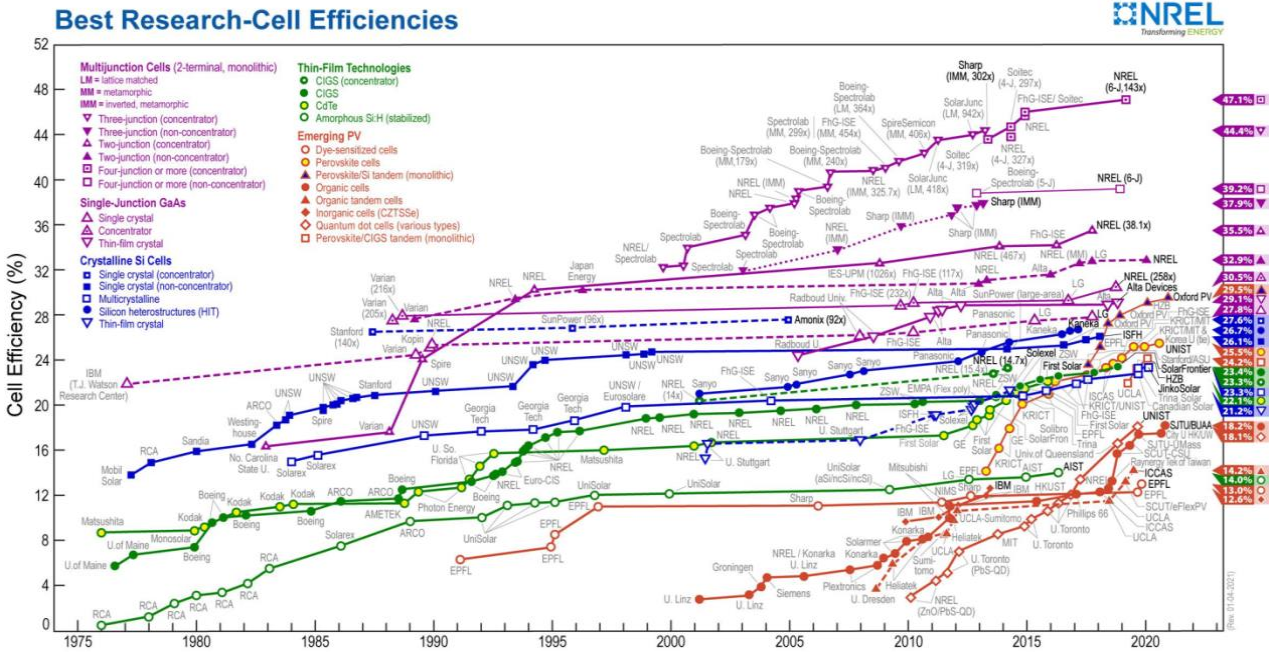

*Figure 5.13 Current progress in solar cell development. US National Renewable Energy Laboratory 2021 https://www.nrel.gov/pv/cell-efficiency.html).*

On rotating worlds, the power of the Sun is periodic, and Solar power must be stored where night-time use is required. Indeed, close to the Sun, solar generators and rechargeable batteries are generally sufficient. Current progress in solar cell development is rapid, as is shown in figure 5.13. Moreover, the next few years will see big improvements in power/mass ratio of solar generators, and energy/mass ratio of secondary batteries. Table 5. 3 lists current and anticipated storage densities available:

| | |
|---|---|
| 50 W/kg | Currently flying on GEO a communications satellite |
| 150 W/kg | With new thin film and flexible technologies |
| 170 Wh/kg | Today's space Li-ion cells |
| 220 Wh/kg | Already in the pipeline (< 5 years) |
| 350 W/kg | Anticipated input from terrestrial R&D growth in Li-ion batteries. |

*Table 5.3. Projected progress for battery technology.*

Even higher storage densities may be possible using new chemistries, for example lithium-sulphur. However, all are eventually limited by the strength of chemical bonds; with bond strengths not exceeding a few eV and while energy storage densities of currently reach $10^3$ Wh/kg easily, $10^4$ Wh/kg would seem to represent a limit imposed by physics The energy stored in fuel cells is bounded by the same chemical limitations.

Table 5.4 sumarises the approximate solar power available relative to that at Earth orbit:

| Planet | Solar power / Earth | Wm$^{-2}$ | Comment |
|---|---|---|---|
| Mars | 25% | 312.5 | Note atmospheric dust storms can reduce power to 10% of Earth |

| | | |
|---|---|---|
| Jupiter | 4% | 50 |
| Saturn | 1% | 12.5 |
| Uranus | 0.25% | 3 |
| Neptune | 0.1% | 1 |

*Table 5.4 Approximate solar power available for selected planets, shown relative to Earth (column 2) and in Wm$^{-2}$. (column 3). The figures are at the top of the atmosphere. Note that the atmosphere of Mars can sometimeshave a high opacity due to suspended dust.*

However, the solar panel area of 64 m$^2$ .for the ESA JUICE mission to Jupiter seem to suggest a realistic engineering upper limit. Solar power efficiencies have been pushed to ~30%, and thus different technology is needed for Saturn and is certainly essential beyond.

By contrast, an RTG provides 10$^5$ Wh/kg and a space nuclear reactor 10$^6$. Wh/kg. Since current High quality solar cells for space use already exceed 30% (eg US National Renewable Energy Laboratory *https://www.nrel.gov/pv/cell-efficiency.html*), it is clear that beyond Jupiter technological improvements to cells and batteries can never be sufficient to power space missions with present day capabilities. Currnetly Currently nuclear power appears to be the only option.

### 4.4.3. Nuclear Power in space

Two main nuclear power system (NPS) technologies are considered for use in space:

1) Isotopic Power Sources are used for heat generation (RHU) and/or for power generation, (typically less than 1 KW) via Radioisotope Thermoelectric Generator (RTG) generally using Stirling engine systems;

2) Nuclear Reactor Systems (NRS) can be used for both heat and power generation and generally provide far more power than RTG. They are however, intrinsically heavy, provide background radiation which can swamp measurement systems, and pose considerable safety issues.

The main applications for NPS in space fall into three classes of application:

- Planetary landers and rovers (RHU)
- Outer Solar System and Deep Space Exploration (RHU and RTG )
- Nuclear Electric Propulsion (electric power from NRS and or RTG)

### 4.4.4. Isotopes suitable for Space Use

In terrestrial applications, where the mass of shielding is not so problematic, 90Sr has been the favoured isotope for RTGs. A quantified assessment of shielding requirements in the context of modern radiological safety standards is necessary to establish if 90Sr can be used in space with a useful power-to-mass ratio. Those isotopes potentially suitable for spacecraft decay-heat nuclear power systems are listed in Table 5.15.

| Isotope | Half-life | Specific Power. ($W_t$/g in elemental form) | Decay Mode and Radiative Emissions | Suitable Chemical Form(s) | Production Route and availability. |
|---|---|---|---|---|---|
| $^{238}$Pu | 88 years | 0.568 | α. Some neutrons from spon-fiss and (α,n). | $PuO_2$ $Pu_2C_3$ | Reactor irradiation of $^{237}$Np or $^{241}$Am. World supplies virtually exhausted. |
| $^{241}$Am | 433 years | 0.115 | α. Some neutrons from spon-fiss and (α,n). Significant soft γ. | $AmO_2$ | Chemical separation from aged reactor-grade Pu. (Carried out in civil reprocessing operations in France.) |
| $^{90}$Sr | 29 years | 0.935 | β. Plus bremsstrahlung photons from β shielding. | $SrTiO_3$ $SrO$ $SrZrO_3$ | Fission product –Could potentially be obtained from a nuclear reprocessing plant. |
| $^{210}$Po | 138 days | 144 | α. Some neutrons from (α,n). | HgPo PbPo Po | Reactor irradiation of $^{209}$Bi. Availability is very unlikely. Included in this table only because of previous space use*. |
| $^{144}$Ce | 285 days | 25.5 | β. Also γ, mainly at 134keV. Bremsstrahlung photons from β shielding. | CeO | Fission product –Could potentially be obtained from a nuclear reprocessing plant. |
| $^{3}$H | 12.3 years | 0.326 elemental 0.098 as LiH | β. Plus bremsstrahlung photons from β shielding. | LiH | Reactor irradiation of $^{6}$Li. |

*Table 5.5. Isotope potentially suitable for spacecraft power systems based on radioactive heat from radioactive decay.*

238Pu is technically the best radioisotope heat source for anything other than very short space missions. However, 238Pu forms only a few per cent of standard reactor-grade plutonium, and cannot be separated. Therefore, it must be specially manufactured by irradiation of other actinides (normally 237Np) in nuclear reactors, followed by chemical purification. This means it is very expensive and the worldwide supply is extremely limited. The USA has stated an intention to recommence 238Pu manufacture in the future, but has always refused to make this material available for export. It is understood that production of 238Pu in Russia has also ended, and existing stocks are limited.

Thus although currently nearly all RTG aplications use Plutonium 238, the Pu-238 production process is multi-stage and includes reactor irradiation, is extremely expensive and poses significant safety and security issues.

ESA has produced a rough-order-of-magnitude cost figure for producing PU-238 in Europe for space units in Europe, and found it was beyond realistic affordability limits. ExoMars will use Russian RHUs for thermal control within both the rover and the static lander module, and as such, will be the first European led (and launched) mission to use NPS. As RHUs are the simplest form of space nuclear power system, and contain only small amounts of radioisotopic fuel, the ExoMars approach is a first step for Europe, and opens the debate on whether more complex radioisotopic power generators could be used on future European missions.

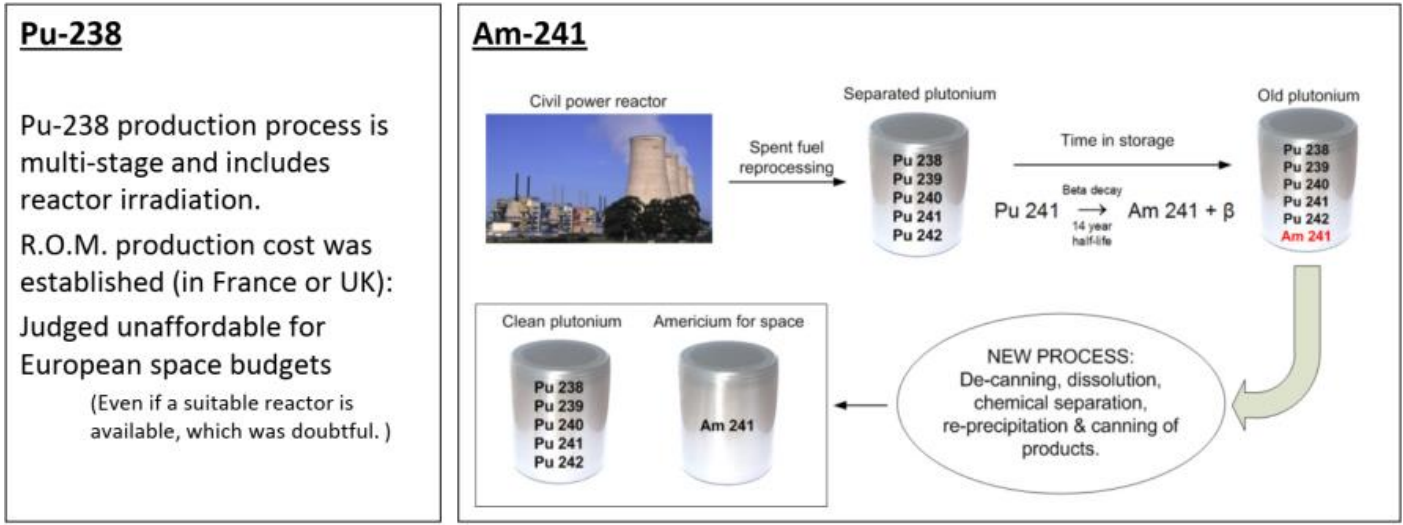

*Figure 5.14. Comparison of Pu-238 and Am-241 isotope production regimes*

An alternative might be a new technology based on Am-241. Americium-241 has never been employed as a radioisotopic heat source, probably because it has only 20% - 25% of the power output of Plutonium-238 (depending on the specifics of compounds, purities, etc.) However, it is an alpha-decaying isotope with only a minor low-energy gamma output, and may be available within Europe as an unwanted by-product of the legacy nuclear fuel reprocessing cycle. Following the chemical separation of mixed plutonium isotopes from the other components of spent nuclear fuel, the plutonium is usually stored in the form of $PuO_2$. During storage, the $^{241}$Pu decays, with a half-life of 14 years, to $^{241}$Am. Hence, 241 is said to "grow in" to the plutonium. Chemical processing techniques can subsequently be used to extract the americium, in a purification cycle. Civil power reacfor waste is separated to produce clean Plutonium and Americium for space use

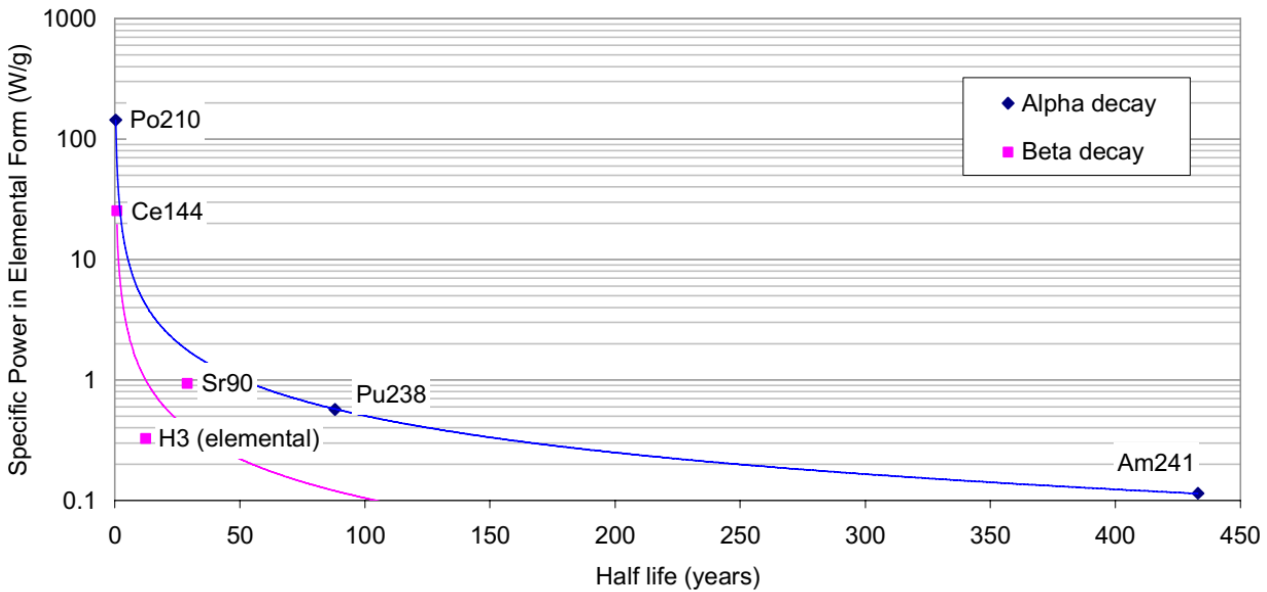

*Figure 5.15. Power vs Half Life for isotopes potentially useful in space. Note thespecific power for Am-241 is around one quarter of that from Pu-238, although the activity is lower.*

Given the severely restricted availability of $^{238}$Pu, the use of 24'Am may warrant further investigation, particularly if an autonomous European NPS capability is required, and $^{238}$Pu remains unavailable or prohibitively expensive. This is particularly the case for out Solar System missions. Although the power available is considerably lower than $^{238}$Pu, it is worth noting that historically outer solar systems based on $^{238}$PU have been relatively profligate with power since it was not an issue. However, of course, the further from the Sun and Earth, the more severe the power issue for telemetry and thermal issues.

### 4.4.5. Conclusions

Power provision currently represents a major bottleneck to future deep space and long duration missions. It is clear that a technological breakthrough of some kind, or a very different political situation to that currently pertaining, will be essential to provide the solutions needed.

## 4.5. Deep space navigation and communications for the future

### 4.5.1. Introduction

Telecommunications with planetary exploration missions is essential in order to send commands and updates to spacecraft so that they can make the measurements required and then transmit the data from which all discoveries subsequently flow. Navigation of planetary missions enables spacecraft to reach their intended targets.

### 4.5.2. Representative Missions

We take two classes of missions as being representative and the most demanding—outer planet systems and missions with small spacecrafts. The targets and objectives can be varied, ranging from conducting a tour of an outer planet system (e.g., *Cassini*, future Ice Giant Mission concept) to detailed studies of particular bodies (e.g., Jupiter Icy Moons Explorer [JUICE], Dragonfly at Titan, Janus concept to explore two asteroids), but a shared technology challenge for both classes of missions is that they have limited power and often are mass-limited.

### 4.5.3. Enabling technologies: current status and future capacities needed

We consider specific improvements on both sides of the telecommunication links: spacecraft and ground systems

**Spacecraft antennas**: Larger antennas enable higher data rates by focusing more of the power on the ground systems at the Earth; all other things being equal, increasing the antenna diameter by a factor of three increases the data rate by a factor of nine. The current standard for outer planet spacecraft is 2 m-diameter, monolithic antennas operating in the X band of the radio spectrum (~ 8 GHz), but with capability at Ka band (~ 32 GHz) for Radio Science. The Mars Cubesat One (MarCO) spacecraft carried 0.5 m-scale, deployable antennas. For small spacecraft, deployable antennas are at the 1 m scale and poised to become larger in the near future (~ 3 m). For outer planet missions, it may be possible to adopt 10 m-diameter, deployable antennas that have been developed for satellites in Earth orbit. However, these antennas would need testing under the environmental conditions experienced in the outer Solar System (most specifically, the low temperatures) to ensure that they remain structurally sound. For both outer planet systems spacecraft and small spacecraft, adopting larger antennas may require improvements in spacecraft pointing.

**Frequency**: As noted above, the current standard for deep space telecommunications is X band. However, switching to Ka band also enables more focused transmissions and therefore higher data rates. All other things being equal, the factor four ratio of frequencies between Ka- and X bands implies a factor 16 improvement in data rate; in practice, lower but still considerable increases in data rate are obtained. As with larger antenna, the use of Ka band likely requires improvements in spacecraft pointing. Of course, this switch is only useful if both spacecraft and ground systems adopt Ka band, but both NASA's Deep Space Network (DSN) and ESA's Deep Space Antennas (DSA) have multiple antennas in their networks capable of receiving at Ka band.

**Power Efficiency**: Current systems employ travelling-wave tube amplifiers (TWTAs) that, while being 50% efficient, are also somewhat fragile and large. Considerable development of gallium nitride (GaN) solid-state amplifiers has occurred for terrestrial applications, and GaN amplifiers have the promise of being both more robust and smaller than TWTAs. However, GaN amplifiers do not yet have efficiencies comparable to those of TWTAs, and technology development of GaN amplifiers with efficiencies comparable to or better than those of TWTAs could yield benefits across all classes of missions.

**Ground Systems**: In order to receive the faint signals sent by distant probes, large antennas (34 m, 35 m, and 70 m-diameters) equipped with cryogenic microwave receivers are used. The cryogenic microwave receivers are already operating near the quantum limits; as noted above, both the NASA Deep Space Network (DSN) and the ESA Deep Space Antennas (DSAs) have multiple antennas equipped for Ka band; additional antennas are being constructed for both the DSN and the DSAs. Additional antennas yield benefits for all missions, but will be useful most particularly as many more small spacecraft are likely to fly over the next few decades. For outer planet systems and outer Solar System missions, multiple antennas can be arrayed together, yielding additional improvements in received power and higher data rates, a technique that was used repeatedly for the Voyager and New Horizons missions. A potential path forward would be to continue adding antennas and considering somewhat smaller diameter antennas to allow for flexibility in handling small spacecraft and arraying for more missions.

Deep space navigation techniques have already enabled small spacecraft to travel the inner Solar System (MarCO) and conduct dramatic flybys in the outer Solar System (Voyager encounters with Uranus and Neptune, *Cassini* flyby of Enceladus, New Horizons at Pluto-Charon). Two, somewhat related areas of improvement are feasible, though. :

**Increased Clock Precision**: Navigation and time keeping have a long history. The current standard deep space clock is an ultra-stable oscillator (USO), achieving an Allan deviation of $10^{-13}$ over 1000 s. Atomic ion clocks, such as the Deep Space Atomic Clock (DSAC), currently in flight in a technology demonstration mission, can achieve as much as a two order of magnitude improvement in precision. This increased precision yields not only improved precision in orbit determination, but also enables longer duration "unattended" or autonomous operations by the spacecraft.

**Autonomous Navigation**: Through the use of "beacons," spacecraft can determine their orbits autonomously. These beacons can be stars, asteroids with well-determined orbits, or, in outer planet systems, the moons of the planet itself. In all cases, precise astrometry is needed, but ESA's Gaia mission is delivering vast improvements in the astrometric precision for all of these "beacons." Autonomous navigation is required in the outer Solar System due to the long light travel times, and it may become increasingly important for small spacecraft as their numbers increase. The use of autonomous navigation techniques has already been demonstrated in a limited sense on the Deep Space 1, but longer duration demonstrations would be beneficial.

## 4.5.4. Main conclusions and suggestions concerning future developments

In many cases, these various technical elements of deep space telecommunications and navigation are being developed or demonstrated. Between now and 2061, there would be ample opportunities to infuse these new technologies into missions.

## 4.6. Autonomy control and health management

## 4.6.1. Introduction

A close partnership between people and semi-autonomous machines has enabled decades of space exploration; but to realize the Horizon 2061 representative missions, future systems must be increasingly autonomous, enabling scientific exploration of regions that have thus far remained inaccessible. Increased autonomy also improves the quality and yield of science data, by allowing better and more reliable utilization of observing time, capturing unpredictable exogenous events of interest, and classifying and prioritizing on-board data, resulting in better use of limited downlink resources.

Autonomy is fundamentally about moving decision-making capabilities to our machine proxies. This transfer of responsibility changes the nature of the exchange of information between humans and autonomous systems. It is also core to solving the explainability and trust challenges with highly autonomous systems. Regardless of the type of autonomy we are planning to deploy, there is a common issue of determining the proper communication between elements in these systems, and between these systems and their operators. The unambiguous transfer of intent from operators to autonomous systems, and the ability of these systems to explain the choices they made, are the core of building trusted autonomous systems.

## 4.6.2. Requirements from representative missions

Autonomy is a tool that both enhances and enables aspects of science missions. In particular, application of autonomy is needed when the uncertainties associated with execution of an *a priori* plan are high. When uncertainties are low, *a priori* plans can be developed and applied with high confidence that they will complete as expected. However, when the uncertainty of plan execution increases – due usually to dynamicism in the environment or system, and compounded by limited communication opportunities and bandwidth – the goals, objectives and policies must be transferred to the system in order to assure high confidence in achieving mission success.

A few representative missions where autonomy is a critical technology, and the envisioned application of autonomy are described below:

- **Future Giant Space Observatories:** Missions such as HabEx would apply autonomy and heath management to preserving observing lifetime, preventing interruptions to observing, maintaining safety, reducing necessary downlink bandwidth, and for control of the starshade.
- **Earth-Moon System Missions:** The proposed NASA Intrepid rover requires increased levels of autonomy to meet the science goals of the mission (driving 1800 km in 4 years). This requires advances in autonomous surface navigation, to plan and safely execute drives, and health management to maintain driving capability and survival during the lunar night.
- **Terrestrial Planet Missions:** Proposed missions such as the NASA/ESA Mars Sample Return mission and the Venus flagship mission will apply autonomy in a variety of ways. Orbiting assets must be coordinated, including autonomous rendezvous for sample-return missions and *in situ* assets must be capable of performing science in multiple locations, maintaining safety during travel, and collecting science in time- and resource-constrained situations.
- **Giant Planet Systems:** Missions to the Ice Giants (Uranus, Neptune) and Ocean Worlds (Enceladus, Europa and more) will employ autonomy to mitigate long communication delays, bandwidth constraints and episodic contact with operations, while maintaining long term reliability, and enabling operation of surface explorers despite communications delays due to increased light travel time.
- **Small Bodies:** Missions to interstellar visitors and long-period comets will face challenges due to incomplete knowledge of the body, high relative velocity at encounter and long one-way light-time in communicating with Earth. In addition, the single and time-constrained opportunity to make observations makes it essential for the planning and health management functions to autonomously adapt the observing plan to account for new information about the body, or failures or degradations in the observing platform.
- **Heliosphere, Solar System, Interstellar Missions and Beyond:** An interstellar mission is perhaps the greatest autonomy challenge, given the limits in communication capabilities and lack of knowledge about the environment. Such a mission will need to be exceptionally self-sufficient and self-directed – able to respond to circumstances that cannot be anticipated and continue to satisfy mission objectives despite the unavoidable degradations of long-lifetime missions.

While there are mission-unique aspects to the deployment of autonomy, there are a set of common technology challenges in the deployment of increasingly autonomous systems. These include:

- Insufficient systems engineering methods and tools limit our ability to design and validate autonomous systems, resulting in time-consuming development;
- Issues with explainability and trust, limiting operational use;
- Lack of an architectural framework, which impedes the integration of state-of-the-art control and machine reasoning technologies;
- Unavailability of high-performance, fault-tolerant, space-rated computing platforms, which restrict the set of technologies that could otherwise be leveraged.

## 4.6.3. Current status and future capacities needed

To reach an enabling level, the community must reach a state where it is straightforward to deploy systems with adequate trust in their capabilities. This will be enabled by well-characterized algorithms and system behavior, and supported by development, verification/validation and operations processes and tools. These systems will have the ability to adapt to circumstance, always performing safely, completing objectives despite circumstances, and relying on operator support when not.

The NASA Autonomous Systems Capability Leadership Team organizes autonomy technology into the following categories:

- **Situational and self-awareness:** Interrogation, identification and evaluation of both the state of the environment and the state of the system;
- **Reasoning and acting:** Analysis and evaluation of situations for decision making;
- **Collaboration and interaction:** Two or more elements or systems working together to achieve a defined outcome;
- **Engineering and Integrity:** Design considerations, processes and properties necessary to implement autonomy.

Work is progressing along all these areas, with the following examples of capabilities that are currently available: On-board planners (M2020), ground-based planners (ROSETTA), distributed control approaches (ERGO), terrain classification (M2020), Risk- and resource aware control (MIT), and Multi-agent mobility control (CARACaS).

Continued technology development provides a rich field of solutions and capabilities to advance the use of autonomy. Advances in goal-directed operation, model-based reasoning, and situational awareness allow operators and scientists to focus on objectives and oversight, while the deployed system determines how to safely perform its assigned objectives. Progress in development of systems engineering processes, system and environmental models and formal behavior specifications enable more rigorous analysis and "correct-by-construction" design specifications, leading to guarantees of system behavior.

Further, advances in artificial intelligences and machine learning enable on-board learning and model adaptation that allow autonomous systems to adjust to conditions over the duration of the science mission. New, promising avenues of technology development include deep learning systems, physics-based reinforcement learning, information-maximizing measurement strategies, and techniques that enable adaptation and evolution of behavioral models. In addition, engineering practices, especially synthesis and analysis functions, driven by requirements that specify acceptable behavior, will need to evolve in order to develop and certify systems using these technologies. To meet the needs of the most challenging future missions, advances and additional technology solutions in promising areas such as general problem solving, automated means-ends analysis and automated hypothesis generation and checking, will need to be explored.

### 4.6.4. Main conclusions and suggestions concerning future developments

Advances in autonomous systems technology will dramatically increase science return by extending the reach, productivity, and robustness of future science missions. This includes capabilities to recognize and exploit opportunities for unique science observations, better approaches to objectives, pruning of low consequence efforts, improved resource utilization, early hazard avoidance, and integrated health management. For missions of the future, *a priori* planning and completely reactive behaviors are not generally sufficient. Instead, systems will need the ability to foresee future possibilities and to relate them to current plans and objectives.

Future developments in this technology areas should include:

- Architectural standards and patterns to enable integration and deployment of state-of-the-art control and machine reasoning technologies, and corresponding advances in engineering processes and tools;
- Assessments and guarantees of system behavior, enabled by principled design techniques and advancements in simulation and formal methods;
- Distributed, model-based control, with control elements organized by frequency and scope, and enhancement of these techniques with data-driven technologies such as machine learning;
- Approaches for plan expressiveness and flexibility to allow effective transfer of intent and explanation between control elements;
- Integration of health management functions into the core on-board planning and execution loop, including off-nominal circumstances in the scope of these functions.

# 5. Science platforms

## 5.1. Introduction

The diversity of types of scientific platforms for deep space exploration missions is expected to expand spectacularly by 2061, as these platforms will need to be adapted to an ever increasing number of planetary environments. For all types of exploration missions, landing on a solid or liquid surface or exploring an atmosphere environment, Entry, Descent, Landing and Ascent (EDLA) technology is a critical enabling element.

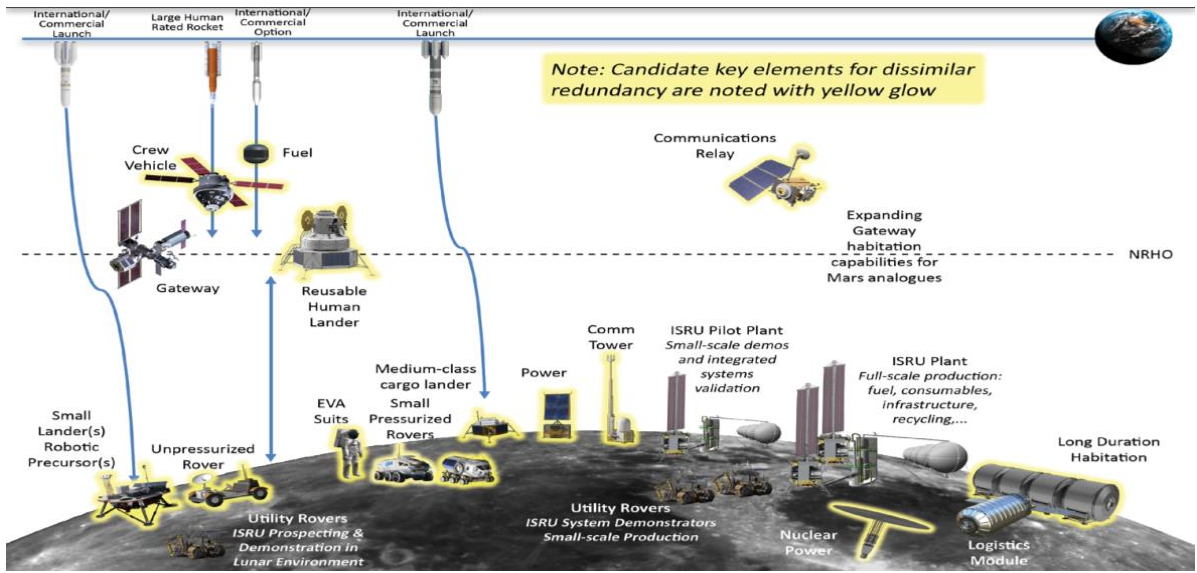

*Figure 5.15. Infrastructure and mobile platforms for sustainable lunar activities (Credit：NASA）*

For science missions using a human lunar/Mars base, as shown in Figure 5.15, what is covered by "science platforms" includes both surface mobile equipment and fixed infrastructure. Planetary surface mobile elements include small cargo landers, pressurized or unpressurized rovers, sustainable power platforms, long duration habitation, In-Situ Resources Utilisation (ISRU) Pilot plant, communication tower... Robotic exploration agents, smart instruments, and human robot teaming also become very important.

## 5.2. Small Spacecraft Orbit Insertion

A major challenge in planning small spacecraft missions is the method of their insertion into orbits around solar system bodies. Small spacecraft could be carried to orbit by larger orbiters if suitable attachment and release systems are available. Placing conventional propulsion systems onto small spacecraft has thus far not been practical and more suitable systems are needed. This challenge needs to be solved through a dedicated technology or alternative insertion techniques in order to achieve the full potential of small spacecraft, for instance for Radio Science experiments (see section 2.2.3). Aerocapture and aerobraking technologies for small spacecraft can enable them to achieve orbit of some bodies as discussed in Austin et al. (2020).

## 5.3. Entry, Descent, Landing and Ascent (EDLA) technologies

### 5.3.1. Introduction

EDLA technology developments, conducted in a coordinated and sustained manner, are key elements enabling a broad range of future missions which includes robotic deep space missions (e.g, Mars, Venus, giant planet moons, small bodies landing or sample return mission) and human Lunar/Mars exploration missions. Different types of EDLA challenges are found with the different planetary environments. Return to Earth with different velocities from these different destinations will also involve EDLA techniques.

### 5.3.2. Requirements induced by representative missions

The different representative missions requiring EDLA can be taken from chapter 4 (Pillar2 report):

1) **For the Earth-Moon System**, possible missions include miniaturized, low-cost lunar landings for ISRU and large human lunar landings missions by 2040, first demos of ISRU life support missions and the deployment of the international lunar village by 2041-2061;

2) **For Terrestrial planets**, possible missions include the Mars sample return mission by 2040, Venus sample return mission and possibly a first human mission to Mars during the 2041-2061 period.

3) **For Giant planets systems**, possible missions include Gas giants moons landers or subsurface explorers, Flagship missions to Uranus or Neptune including atmospheric entry probes by 2040. Gas giants moons and rings subsurface explorers, Europa or Enceladus Plumes and subsurface sample return missions are foreseen for 2041-2061.

4) **For small bodies**, possible missions include comet and Trojan sample return by 2040; application of ISRU to asteroids, and possibly a first human mission to a NEO are foreseen for 2041-2061.

Different missions require different types and mass scales of landers. They can be divided into human lander and robotic landers, and into large, medium and small landers according to their mass and external dimensions. The different extreme environments of different missions induce different requirements on EDLA technology. The resulting relationship between these different requirements and the lander is shown in Table 5.5.

| Missions type and Extreme environment | Robotic Sample return | Human exploration | Planetary atmosphere | Moons or rings lander or subsurface exploration | Typical Extreme environment |
|---|---|---|---|---|---|
| **Earth-Moon System** | small-Scale lander for Moon | Heavy Scale lander for Human Moon mission | | | - Earth Return velocity (11-13km/s) |
| **Terrestrial planets** | small-Scale lander for Mars or Venus | Heavy Scale lander for Human Mars mission | small-Scale lander for Mars or Venus atmosphere | small-Scale lander for Mars and its Moons | - >8km/s at Mars, >12km/s for Earth Return<br>- High pressure and high temperature for Venus<br>- Sulfuric acid clouds for Venus<br>- Dust storm environment for Mars |
| **Giant planets systems** | Small/Middle-Scale lander for Europa or Enceladus subsurface &plumes | | Middle-Scale lander for Gas giants moons atmosphere | Small-Scale lander for Gas giants moons & rings subsurface | - Earth Return velocity(17-20km/s)<br>- Low temperature<br>- High radiation environment<br>- Corrosive liquid or ice environment |
| **Small bodies** | small-Scale lander for Comets and Trojans | Heavy Scale lander for Human NEO asteroid mission | | | - Earth Return velocity (11-20km/s)<br>- Dust or ice environment |

*Table 5.5. Relationship between different mission requirements and the lander. Small Scale lander: Mass usually 0-2T, especially for robotic sample return mission. Middle Scale lander: Mass usually 2-10T, especially for Giant planets lander mission. Heavy Scale lander: Mass usually 10-60T, especially for human lander missions.*

### 5.3.3. Current status and future capacities needed

Based on the type of EDLA capacity needed, listed in table 5.5, one can determine the enabling techniques required, as shown in figure 5.16.

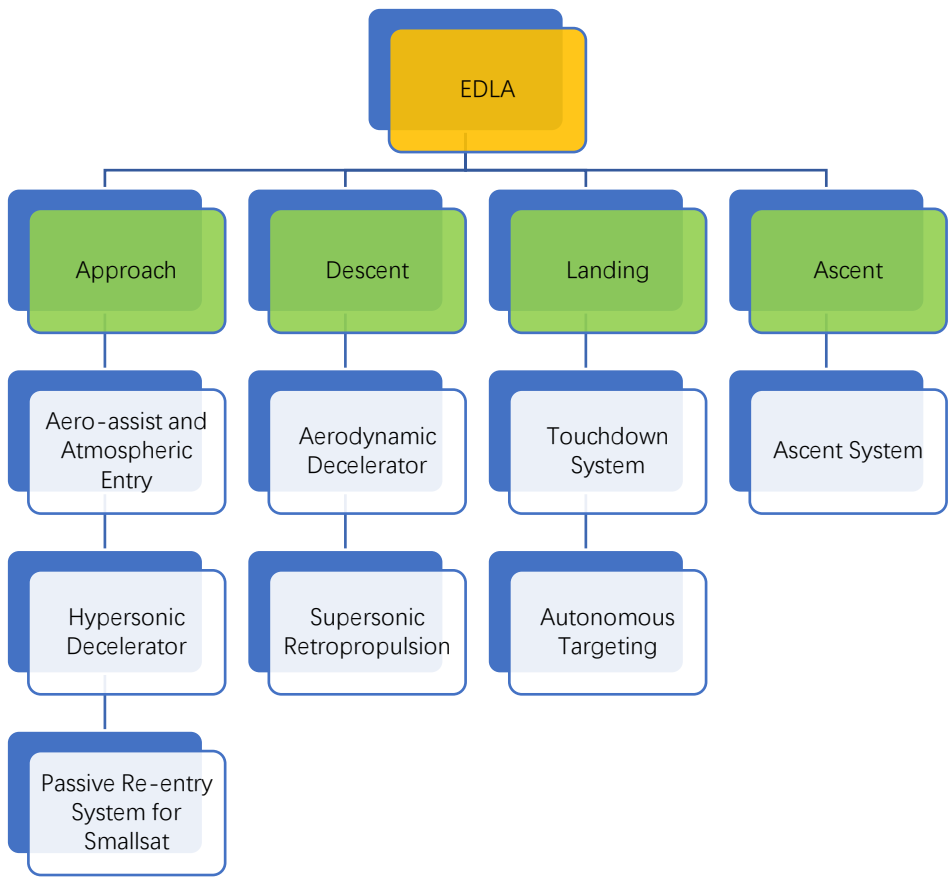

*Figure 5.16. The architecture of EDLA technology*

## 5.3.4. Conclusions and suggestions concerning future developments

Despite the different requirement of different missions, the key performance characteristics of EDLA technology developments are landed mass, reliability, cost, landing site elevation, and landing accuracy. Like EDLA subsystems, these characteristics interact with each other. Reliability results from thorough testing and analyses of component technologies, such as thermal protection systems, deployable decelerators, landing hazard tolerance, and separation systems. Reliability might also be improved by increasing the duration of controlled descent as a result of larger drag devices applied earlier, technology development for precision landing (reliant on detailed site information for a priori hazard identification), hazard avoidance, and mitigation of site hazards created by terminal descent propulsion. For missions like Mars sample return, planetary protection requirements place higher emphasis on robustness and reliability. Low cost is enabled by improved simulation and ground-to-flight extrapolation, and by incorporating high-g landed systems into mission architectures where applicable.

On the other hand, lower-g entry systems, such as deployables, can enable sensitive science instrument and human delivery, thus providing new and exciting science and exploration opportunities. Delivered mass can be increased or enabled by using more capable Thermal Protection System (TPS) for the more challenging environments, larger entry vehicles, larger drag and/or lift devices applied at higher speeds and altitudes, descent phase (supersonic) retro-propulsion, and more efficient terminal descent propulsion. Landing site access can be increased by using a TPS that permits higher entry speeds (allowing a wider range of targets), small body proximity operations, increased altitude performance by increasing drag early in the descent, and increased trajectory range and cross-range with higher precision, allowing a wider range of safe sites. Higher control authority, particularly in the case of large deployable systems, also enables higher precision in the entry phase. Both precision landing and hazard avoidance are enabled by a combination of more advanced terrain sensing and algorithms with more capable terminal descent propulsion and guidance to divert the lander to the desired target. All of these objectives benefit from improved modeling of the systems and of the natural environments.

So, for missions such as Robotic Sample return, Human exploration, atmospheric entry probes, Moons or rings landers or subsurface exploration, EDLA is a key technology to secure success.

## 5.4. Planetary surface mobile elements

### 5.4.1. Introduction

Mobility will be the main issue for the future exploration of planetary bodies. This mobility will be achieved by two different types of platforms: rover and drone (or Unmanned Aerial Vehicle: UAV). The first will explore the planetary surface by moving on the surface, whereas the drone will be able to fly at different altitudes over the surface. Both provide different exploration scenarios in term of mobility and payloads. In both cases, the main driving factors governing the manufacturing and use of the platforms are the travel range and the transport capability of payloads (in term of mass and volume).

### 5.4.2. Requirements induced by representative missions

Rovers are used for the exploration of the Moon and Mars and they have dramatically improved in term of autonomy, driving capabilities, and travel ranges. The number of instruments has also increased up to Curiosity which, using an arm, has been able to operate a number of proximity sensors. However, the rover operations remain mostly limited by the difficulties in avoid obstacle and in performing long drives. Opportunity has been able to drive autonomously for 1250m during an almost 3000m drive. However, the scenarios for the post-MSR exploration involve long range mobility with a number of instruments and with the capability to negotiate treacherous terrains. These will be precursor rovers which collect data for the planning of the first human Martian missions, and for the selection of landing sites, and so should be able to drive autonomously over large ranges. Once the human exploration is begun the robotic components of operations will be extremely important in mission planning, and supporting astronauts during Extra Vehicular Activities (EVAs). Careful planning of human operations can reduce the exposure time during EVAs and a rover following and supporting human crews will be an important safety element.

The exploration of the Moon will follow, or perhaps predate, a very similar strategy to Martian one. Probably a robotic component will be mostly used during the external operations outside the habitats. Navigation will be easier due to the simpler geomorphology of the Moon and its more homogeneous surface.

Drone utilisation will rapidly increase, with the development of the roadmap towards the Martian human exploration. Currently only one single drone has flown in the Perseverance mission. However, NASA has selected Dragonfly as a drone mission to Titan. The launch will be in 2026 and the arrival in 2034. There is no doubt, that drones will be useful exploration tools for the atmosphere-bearing planets. Drones exhibit a large and diverse range of utilisation from pathfinder simple control of rover operations to complex geological surveys.

### 5.4.3. Current status and future capacities needed

The technological effort in developing rover surface platforms must aim at two different objectives: (i) the first is to increase driving distance, and the ability to navigate rough surfaces and steep slopes, combined with improving autonomous hazard avoidance. (ii) the second is to improve GNC (Guidance, Navigation and Control) capabilities, using more flexible and human-supervised tools such as virtual reality and augmented reality.

A large effort currently under way spent to achieve autonomous driving, which is a necessary step in order to perform a smooth and safe drive. However, human supervision remains fundamental in (i) planning drives, particularly those that include long distances, scientific targets and complex operations; (ii) driving in treacherous terrains where subtle features can remain unnoticed to AI tools; (iii) navigating and controlling a rover that cannot totally controlled by autonomous GNC.

The Virtual Reality environments can be constructed by high-resolution images and Digital Elevation Models (DEMs) from satellite and drones. The initial data sets can be enriched by images and 3D rendering obtained by the rovers themselves. This system will create a virtual system for the planning of long-range driving and for robust guidance and control of operations.

Technologies are available and tested in Earth settings, and are starting to be used for the selection and certification of landing sites for future missions. Tests for analysis of this approach to rover and drone navigation and control by virtual reality being studied in several laboratories worldwide. The application of this approach to real missions should be ready for implemented shortly.

## 5.5. Robotic exploration agents, smart instruments, and human-robot teaming

### 5.5.1. Introduction

Future space missions will exhibit significantly greater decision-making capability than current missions (Chien and Wagstaff 2017). Missions of all classes will utilize onboard flight software to:

(1) detect science events of interest and respond autonomously to improve science;

(2) handle anomalies more effectively to reduce downtime and science lost;

(3) operate with improved efficiency, with robust handling when execution or resource usage varies from predicted.

In some cases, multiple space assets will extend this coordination across platforms autonomously, either through space networking or via a ground link, allowing autonomous ground coordinated response. Additionally, ground-based automation will enable enhanced monitoring of operations, autonomous responses to anomalies, rapid response modification of operation plans. Enhanced decision support tools to can give ground operations teams greater situational awareness and understanding of operations planning.

This set of technologies can be categorized as follows.

● **Science event detection and onboard data analysis, including summarization**,

includes all manner of onboard analysis of data to determine conclusions from lower level data including data science and machine learning (both supervised and unsupervised). These technologies can be applied to science or engineering data.

● **State estimation, mode identification and recovery, and integrated vehicle health management**

includes all manner of technologies to understand the spacecraft state and manage the health of the vehicle (see related section by Day et al.)

● **Planning, scheduling, and robust execution** includes technologies related to projecting future mission activities and executing them to achieve mission goals.

● **Multi agent coordination** includes all aspects of the above technologies when there are multiple assets with limited/unreliable/delayed communications and the mission-required coordinated activities to achieve goals (Chien et al.2000).

● **Human machine interaction** includes methods for enabling humans to understand the behaviour and state of both the spacecraft (or multiple spacecraft) technologies above, as well as the AI software ("explainable AI").

● **Adaptation, learning, evolution** includes technologies to enable the above reference systems to improve their performance by learning and updating models of the world, their activities, and AI systems.

Give a synthetic description of the technology item which is the subject of this section, and of its generic domains of application and roles in a planetary mission.

## 5.5.2. Requirements from representative missions

All future mission classes can benefit from the AI technologies mentioned above, and indeed numerous identified target missions within the 2061 timeframe are probably not possible without AI/Autonomy technologies.

Surface missions such as future Mars rovers (Gaines et al. 2020) and the proposed Europa Lander Mission Concept (JPL 2020) have intimate robotic interactions with the environment and as such experience significant variability in execution times, activity failures, rates of progress, and resource usage (Gaines et al.2016). Onboard scheduling and flexible execution, such as implemented for the Mars 2020 Perseverance Rover (Rabideau and Benowitz 2017, Chi et al. 2018, Chi et al. 2019, Chi et al. 2019) and proposed for the Europa Lander Mission Concept (Wang et al. 2020) can help these missions to achieve their mission goals by adapting onboard activities to execution feedback, in order to achieve mission goals in a wide range of situations.

A future generation of missions to staggeringly hostile and unpredictable environments such as a Europa Under-Ice Submersible (New Scientist 2017, Wirtz et al. 2016, Branch et al. 2020), an Enceladus Vent Explorer (Ono et al. 2018), and a Titan Aerobot (Hall et al. 2006) or rotorcraft (Lorenz et al. 2018) would all require significant autonomy. More conventionally, orbiters and observatories will greatly benefit from onboard and ground-based automation. Already the Spitzer Space Telescope utilizes onboard autonomy with the Virtual Machine Language (VML) to adapt to execution variations in acquiring guidestars for observations - handling a list of observations for each operations period and accomplishing as many as it can, given execution variability (Peer et al. 2005). Ground-based use of AI scheduling is commonplace in automating observatory scheduling including notably the Hubble Space Telescope (Johnston et al. 1994), Spitzer (Mittman et al. 2013), and plans for the James Webb Space Telescope and many other missions (see (Chien et al.2012)). Flyby missions (Whittenburg 2019) can also benefit tremendously from onboard Artificial Intelligence. AI would enable the flyby spacecraft to detect plumes (Fuchs et al. 2015), surface volatiles, other surface features, or even satellites and re-target to image and study appropriately (Chien et al. 2014, 2016). However great care must be taken that unexpected observations do not compromise science goals, as occurred on the Giotto mission. This approach is also very relevant to "reverse flyby" mission (Castillo-Rogez et al. 2019) such as to observe an interstellar visitor or long period comet where the spacecraft cruises to a rendezvous point to observe an object that flies by at a relative velocity 40 km per second or greater. As with traditional flybys, the high relative velocity requires onboard AI to detect, track and re-target to observe the target effectively.

A proposed mission to fly 100 cubesats or smallsats (Keck 2014) to autonomously survey the Near Earth Object (NEO) population would likewise require extreme autonomy to avoid the cost of operating 100 spacecraft individually.

Even further afield, the Solar Gravity Lens mission (Keck 2017), missions to the Interstellar Medium (Stone et al. 2014, Staehle et al. 2020) or to nearby stars (Freeman et al. 2017) would require even greater autonomy. With mission durations of decades and round-trip light time of many years they clearly require true autonomy.

An orthogonal, but important trend is the increasing use of large-scale constellations. Planet Labs (Planet 2020) and Starlink have approval to launch thousands of satellites. While these are in Earth Orbit, constellations and swarms of spacecraft have many applications to planetary exploration. As these systems have increasingly complex interactions, networking the assets via an Interplanetary Internet (Burleigh et al. 2003) becomes critical. As these systems grow increasingly autonomous, multi-agent AI will become an enabler.

Future concepts which extrapolate these trends include dependent, but agile "daughtercraft" systems deployed from a more capable platform (Karras et al. 2017, Vander Hook, et al, 2019) to explore risky areas such as lava tubes and other types of caves (Whittaker 2012, Troesch et al.

2018), lava tubes . Intelligent cooperative systems would be incredibly useful for in-space assembly of the large aperture telescopes (Lee, et al 2016) or precursor outposts (Huntsberger 2005), which are far too large for single launches. Many such applications for multi agent space missions exist (Rahmani et al. 2019). Furthermore, a network of systems can share or provide AI or data science "services", enabling cost-sharing or multi-institution collaborative missions (Vander Hook et al 2019, 2020). For smaller missions, or where there is immediate access to with more data than can be downlinked to Earth, this can enable significantly more on-site processing and decision making capability.

AI has already been applied to design and evaluation of multi spacecraft radio interferometric missions (Belov et al. 2018, Schaffer et al. 2018), such technologies would be equally useful in operations of such large-scale constellations.

### 5.5.3. Current status and future capacities needed

All of the AI technologies listed above are developing rapidly both in the space and non-space sectors. Below we project some of the most critical technology needs and developments for missions in the 2041 - 2061 timeframe.

● **Onboard science and engineering data analysis.** This area to date has used static techniques (e.g. computer vision (Castano et al. 2007, Estlin et al. 2012,Francis et al. 2017). Additionally, use of machine learning has been carried out off-line, and carefully tracked by ground teams prior to upload [Doggett et al, 2005, Altinok et al. 2017, Wagstaff et al. 2018].

In the future Machine Learning usage will become more prevalent and expand to online learning applications, and additionally use unsupervised learning techniques to deal with more unknown situations (Hayden et al. 2012, Wagstaff et al. 2018). Use of these techniques onboard for automated analysis of engineering data has lagged its use for science data, due to the higher stakes (e.g. spacecraft health and mission loss). This area (especially instrument processing) is a driver of onboard computing needs (see below).

● **State estimation** (see related section);

● **Planning, scheduling, and robust execution** will evolve to become more closely linked to special purpose reasoning. Most current scheduling systems focus on state, resource, and timing problems. While some systems reason about geometry (e.g. coverage) such as for Rosetta (Chien et al. 2015) ECOSTRESS, EMIT, NISAR, (Chien et al. 2019), and spatial aspects, these are less widespread. Other research prototypes have included spatial aspects of planning (Woods et al. 2009, 2014, Gaines et al. 2010, Wettergreen et al. 2014) These technologies will become more mature, and more widespread in usage, including onboard implementation.

● **Multi agent coordination** has not yet been deployed for space applications. However the growing number of multi asset and swarm space applications will lead to increasing deployment of multi-agent autonomy for space, surface, and aerial vehicles, further facilitated by the Interplanetary Internet (Burleigh et al. 2003).

● **Human machine interaction**. As AI systems become more prevalent in space, human machine interaction with these AI systems is of growing importance (indeed identified as a major research topic for the future (Gil and Selman 2019).

● **Enabling technologies**: increases in flight computing enhance many of the AI applications listed, but none more so than in onboard processing of instrument data. Already, instruments such as radar, lidar, and hyperspectral imagers produce data streams of Gbits per second. Analyzing these data streams in near real time opens up a whole range of AI applications, only possible with advances in flight computing. Current efforts to fly more powerful computation include flight of the Qualcomm Snapdragon on the Mars Helicopter (Grip et al. 2019) and flight of the Intel Myriad Chip on FSSCCAT (ESA 2020), however future computing needs for onboard AI will continue to grow (Dally et al. 2020).

As AI comes into the mainstream controlling high value assets such as space missions, Validation and verification of the AI both individually and of the entire system including AI will be a critical area for technology investment and maturation.

The final frontier of AI, intelligent systems, and autonomous systems is for the systems to adapt, learn, and evolve particularly as relating to "extended duration autonomy" of decades in unknown environments beyond frequent human contact (e.g. interstellar missions). Future systems will need to be able to adapt independently in order to survive and thrive for decades.

# 6. Technologies for long-term and sustainable exploration

## 6.1. Introduction

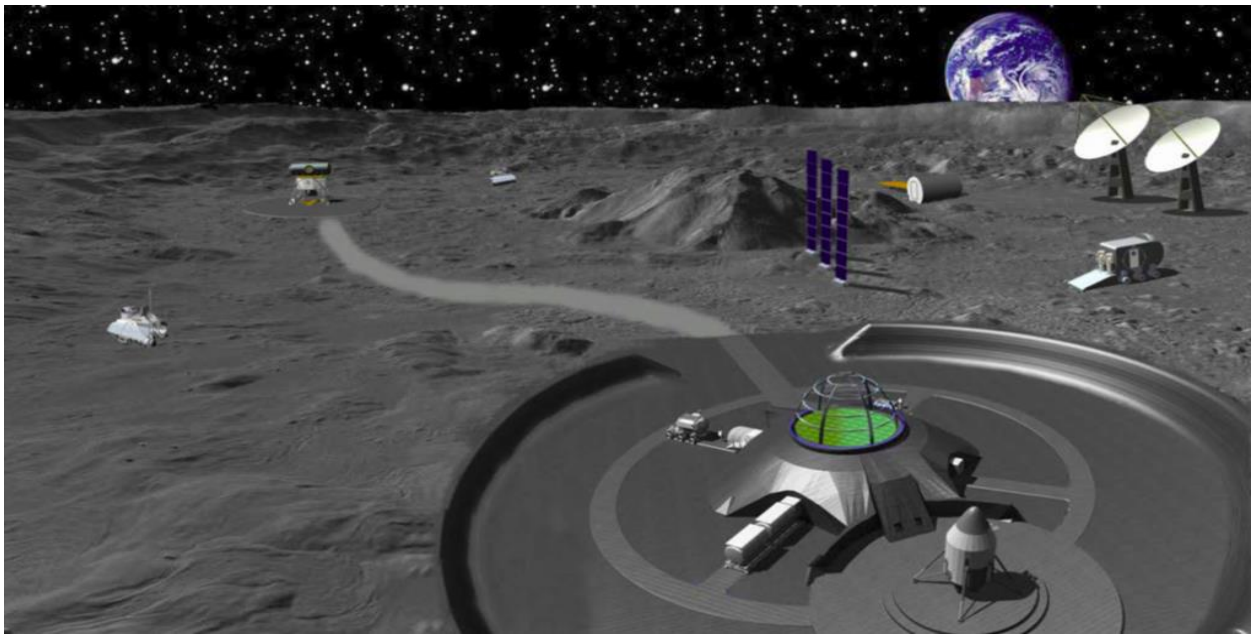

*Figure 5.17. Concept of a manned lunar base presented by CAST*

The Moon and Mars are the next destinations for post-International-Space-Station human exploration, as stated in the NASA Global Exploration Roadmap.

Whether it is for investigation and further usage of the available local resources (volatiles, water, metals) to support sustainable exploration

[https://exploration.esa.int/web/moon/-/61369-esa-space-resources-strategy],

to address science objectives, or to prepare technologies for exploration to further destinations, infrastructures for long-term surface exploration activities are expected to be required.

In order to be able to stay at the destination and have the ability to long-term survival, as well as the ability to return to earth from the destination, we need to build on the ability to live and work at the destination for a long time. The Moon on average is about 360000 km from the Earth, and it is even more difficult to supply a Mars base from Earth, and the launch window from Earth to Mars is only every two years. Therefore, humans must learn to use the resources of the destination to produce oxygen, water, food, building materials in order to solve the problem of long-term survival in the destination.

In addition, in order to be able to reduce launch costs, and reduce the total mass scale of launch vehicles, human also need to learn how to use the resources of Moon or Mars to make liquid hydrogen/ Liquid oxygen and liquid methane/liquid oxygen propellant, and infuse them into Mars ascenders, in order to provide a manned return to the Earth. So, in-situ resource utilization technology must be developed to solve the problem of long-term survival on extraterrestrial bodies.

|  |  | per capita per day/kg | per capita per year/kg | Percentage of total quality/% |
|---|---|---|---|---|
| input |  |  |  |  |
|  | oxygen | 0.83 | 303 | 2.7 |
|  | food | 0.62 | 226 | 2.0 |
|  | Drinking water | 3.56 | 1300 | 11.4 |
|  | Clean water | 26 | 9490 | 83.9 |
|  | Total | 31.0 | 11400 | 100 |
| output |  |  |  |  |
|  | Carbon dioxide | 1.0 | 363 | 3.2 |
|  | Metabolic solid discharge | 0.1 | 36 | 0.3 |
|  | water | 30.3 | 10950 | 96.5 |
|  | Metabolites / urine |  |  | 12.3 |
|  | Sanitary water |  |  | 24.7 |
|  | Cleaning water |  |  | 55.7 |
|  | Other |  |  | 3.6 |
|  |  | 31.0 | 11400 | 100 |

*Table 5.6. Consumables for a lunar base.*

| project | Consumption (t) |
|---|---|
| Consumptive articles for daily necessities | 1.66 |
| Environmental control consumables | 0.81 |
| Drink water | 0.38 |
| Extravehicular consumables | 1.31 |
| Emergency recharge equipment recharge | 1.22 |
| Total | 5.38 |

*Table 5.7. Required consumables for a 90-day cycle on a lunar base with 6 occupants*

In order for human deep space exploration to be sustainable, the ability to assemble and manufacture in space is also required. For large space vehicles (more than tens of tons), it is difficult to send them to the entry point of the Earth-Moon or Earth-Free Transfer Orbit through a launch mission. Therefore, it is necessary to complete in-orbit assembly of large space objects by means of space rendezvous and docking. This also enables production in situ, for example 3D printing a Moon base. In addition to reduce the size of the transportation supply from Earth materials, development of advanced Environmental Control and Life Support (ECLS) system technology, improvements to the closure of the material, and reduction of material also have vital significance for closed planetary vehicles and extravehicular activities. Advanced ECLS technology, also can improve astronaut work efficiency.

Therefore, we generally believe that the development of ISRU in-space assembly and manufacturing, as well as advanced ECLS, are the core technologies that are critical to improving the long-term presence and return capability of human beings on extraterrestrial bodies

## 6.2. In situ resource utilization (ISRU)

### 6.2.1. Introduction

In order to build and manufacture such infrastructure and associated hardware, maximization of the use of material resources available at destination is considered essential. This will allow substantial savings in payload mass, cost and mission complexity, help reduce the dependence on cargo resupply missions from Earth and ultimately increase the sustainability of exploration activities, by supporting the establishment and further expansion of dedicated settlements. This concept is referred to as In-situ Resource Utilisation (ISRU) and has been the subject of renewed and increased research and technology development activities. ISRU is being considered to help address a range of needs for future long term exploration missions, from the extraction of oxygen and water for propellant and life support 0, Starr et al, 2020), to the processing of materials for infrastructure construction and manufacturing(Lim et al, 2017, Naser et al, 2019).

### 6.2.2. Requirements induced by representative missions

Future long-term exploration missions to the Moon and Mars are expected to require infrastructure and supplies to support sustained human presence, as well as robotic and human activities. In-situ resource utilization is considered to be an enabler, to enhance the sustainability, as well as the technical and economic viability of such missions.

The needs for such missions include environmental shielding of the crew against the lunar and Martial environment aspects, among which are radiation, micrometeoroids, vacuum and temperature variations (Anand et al, 2012). Consumables for life support of the crew, including water and oxygen, will also be required.

Protecting the equipment and the crew from the impact of the charged and magnetised abrasive dust will be essential, as this was identified as one of the most significant challenges to lunar surface exploration, during the Apollo missions (Gaier,2007; Taylor,2005). This implies the construction of protective shelters or berms.

Mobility will need to be ensured in an uneven terrain covered with boulders. This includes the need to build landing pads to enable safe landing and ascent of vehicles, as well as terrain levelling and construction of roads, to facilitate rover access.

Energy generation and storage will be fundamental to provide the power required by the various exploration activities.

In-situ manufacturing of hardware and tools for crew activities will also help reduce the dependence on supplies brought for the mission or subsequently delivered from Earth.

The development and demonstration of technologies which allow in-situ utilization of the resources available locally, to address each of the mission needs listed above, will help advance the implementation of sustainable long-term exploration missions.

## 6.2.3. Current status and future capacities needed

The most abundant material resource available on the Lunar and Martian surface is the regolith. i.e. the mixture of dust, soil and broken rock which constitutes the superficial layer of the Moon and Mars (G. Heiken,1991). The regolith is composed of a mixture of minerals which contain oxygen, silicone, various metallic elements and some non-metallic elements.

Several techniques have been investigated to extract oxygen from regolith, by chemical or electrochemical reactions (Schlüter et al,2020). This include recent demonstration that electrolysis could potentially allow extraction of the total oxygen content from the regolith 0. The oxygen extracted by those techniques could be applied for use as a propellant, for life support of the crew or for energy storage in fuel cell applications. The remaining metal-rich mixture (see Figure 5.18) could be separated into individual metals or processed as-is, for construction or hardware manufacturing. The oxygen extraction techniques have been demonstrated so far in the laboratory environment and work to demonstrate them in representative environment is ongoing.

For Martian applications, most processes for oxygen production are based on extraction from the $CO_2$ atmosphere (Starr et al, 2020), including the MOXIE solid electrolysis instrument which has been landed on Mars on the Perseverance rover, as part of the NASA Mars 2020 mission.

Access to and retrieval of water from water ice deposits on the lunar and Martian surface are also envisaged (Shuai Li et al, 2018), supported by scientific observation on water resource availability (Cesaretti et al, 2014). Accessibility to those water deposits, in particular in permanently shadowed regions on the lunar surface, will require development of vehicles and energy sources which allow to operate in those difficult terrains.

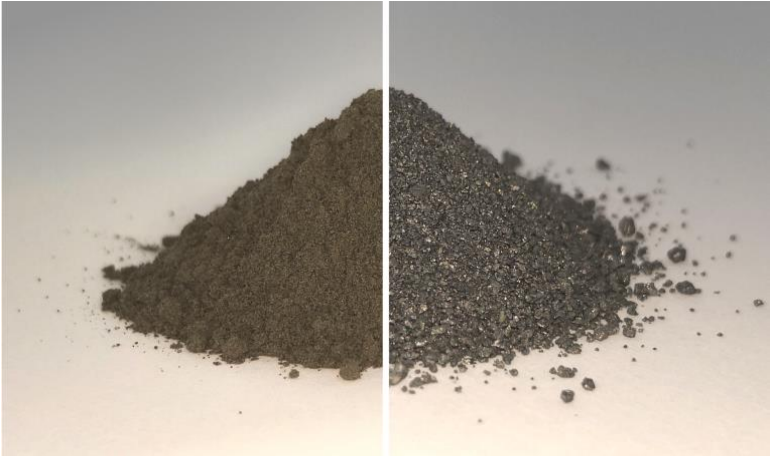

*Figure 5.18. Lunar regolith simulant before (left) and after (right) extraction of oxygen by a molten salt electrolysis process*

The processing of regolith into construction materials and the production of structure demonstrators have been the subject of numerous research and technology development activities (Lim et al, 2017; Naser et al, 2019). Techniques for turning regolith into a solid construction material can broadly be classified into two categories: processes involving binders and processes involving thermal energy. In the first category, the regolith is mixed with a binder which triggers a chemical reaction, leading to the formation of a solid material or a paste-like material which can be extruded. Among the various methods investigated so far, additive manufacturing of three-dimensional demonstrator parts for structural shelters, ranging from a few centimetres to more than a metre, was demonstrated. The techniques used involved spraying a magnesium chloride binder onto successive beds of regolith simulants (Buchner C, et al ,2018) (see Figure 5.19) or mixing the regolith with a phosphoric acid binder (https://www.esa.int/ESA_Multimedia/Images/2018/11/3D-printed_ceramic_parts_made_from_lunar_regolith#.X5cir8s7thw.link), to form a paste which could be extruded into layers.

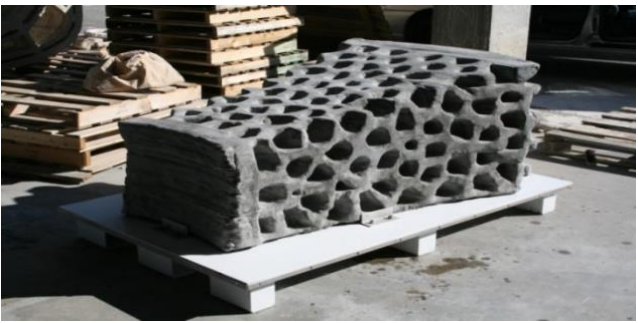

*Figure 5.19. Demonstrator of lunar habitat protective shell segment made by additive manufacturing of lunar regolith simulant, using a magnesium chloride binder*

A lithography-based additive manufacturing technique, used in terrestrial applications for the manufacturing of high-end ceramic parts, has also recently been demonstrated to produce very accurate ceramic tools, using regolith simulant (Karl, D. et al,2020). Examples of manufactured parts are shown in Figure 5.20. This process is limited by the dimensions of the additive manufacturing machine's build chamber. It is therefore more relevant to produce small hardware parts than large infrastructure elements. In addition, post-sintering at temperatures in the range of 1000 °C is required, to consolidate the additively manufactured green body. This requires that sufficient energy is available at the lunar or Martian surface to allow heating to this level of temperature.

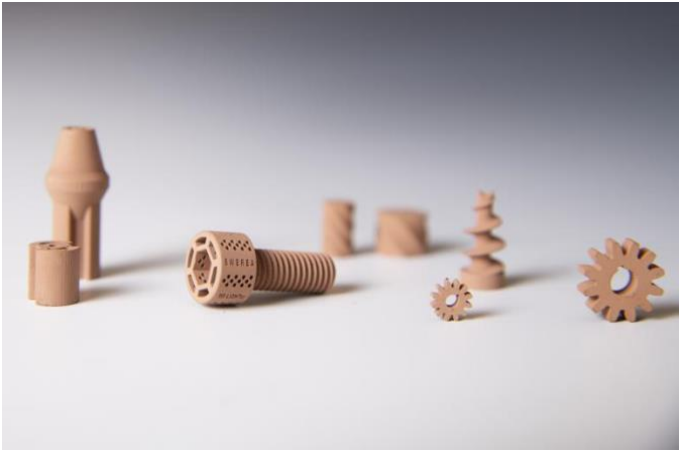

Figure 5.20. Ceramic parts produced by lithography-based additive manufacturing of lunar regolith simulant

Regolith processing methods based on binders require the use of chemicals which are often not readily available on the lunar of Martian surface. They would therefore need to be brought from Earth in large quantities, matching the scale of the intended construction.

More recently, additive manufacturing processes using additives that can be sourced from the lunar or Martian surface have been proposed. This includes using urea as a plasticizer for lunar geopolymers 0 or using water for Martian clay processing(Karl, et al,2020). Such processes help reduce the need for binding agents brought from Earth. However, the availability of the required reagents in sufficient quantities for large scale construction can be subjected to trade-off against other applications which may be considered of higher priority. The use of urea as a nutrient for food production or the sourcing of local water for propellant and life support may be preferred to construction applications.

Technologies involving only the use of thermal energy to consolidate the regolith have been investigated (Lim, et al,2017; Naser et al, 2019). They rely on applying heat to either sinter the regolith – i.e. partially fuse the regolith grains to allow bonding between adjacent grains – or to melt and solidify it. Those thermal processes allow to obtain a consolidated material, which can be used for construction or manufacturing, by using only resources available locally, without the need for additional binding agents.

Various sources of heat have been tested. Additive manufacturing by laser sintering was shown to produce three dimensional parts with highly accurate geometries, but limited to a couple of centimeters in size, due the high power requirements and limited processing diameter of the laser (Fateri, et al.2015).

Studies are ongoing to develop regolith manufacturing processes based on microwave sintering, as this is expected to require lower energy amounts than laser sintering and higher processing speed for large scale construction (Srivastava et al, 2016; Fateri and Cowley, 2019). However, regolith microwave processing tests have been demonstrated so far on specimens of limited dimensions and systems for sintering of two- of three-dimensional areas are yet to be developed.

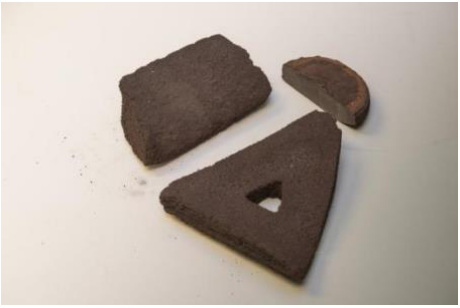

Figure 5.21. Examples of 3D printed parts obtained by additive manufacturing of regolith simulant using concentrated sunlight

In an attempt to use direct concentrated sunlight for heat generation, i.e. without intermediate conversion of electrical energy into laser or microwaves, solar sintering of regolith simulant has been investigated. Systems for solar additive manufacturing using mirrors or lenses for sunlight concentration have been successfully demonstrated, through production of three-dimensional parts of several tens of centimeters in size (Meurisse et al, 2018; Fateri et al, 2019). Figure 5.21 shows examples of such solar sintered lunar regolith simulant parts. Solar sintering appears promising for low-energy, large-scale construction, although the mechanical properties of the consolidated material need to be optimized.

The regolith-based ISRU construction processes could be implemented on the lunar or Martian surface to build habitat or equipment shelters, by fitting relevant printheads on gantry systems or mobile rovers, as depicted in the concept shown in Figure 5.22. Binder- and heat-based processes may also be used to consolidate areas on the lunar or Martian surfaces for landing pads or roads. Implementation of such techniques would require further development, from the current stage of laboratory demonstration, to understand the impact of the extraterrestrial environment – including reduced gravity, thin atmosphere, radiations, thermal environment and charged dust – on the processes and on the manufacturing equipment.

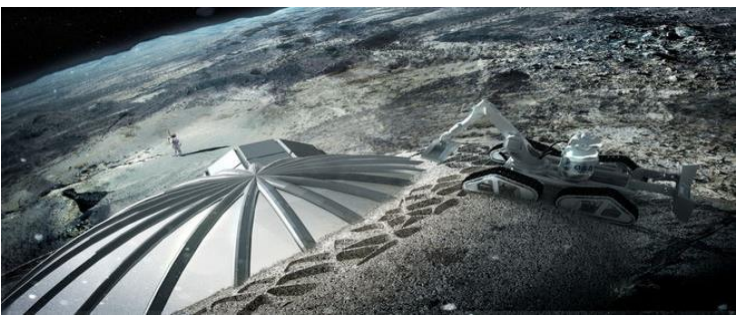

Figure 5.22. Concept for additive manufacturing of a protective shelter from regolith

Another construction application of regolith melting, which is currently being investigated, is the extrusion of molten regolith into fibres, which could then be woven into structural habitat elements by robotic manufacturing (Hanna Läkk, et al, 2018).

For energy-related applications, the potential of regolith processing for thermal energy storage and thermoelectricity generation has also been studied, highlighting the limitations of such an approach. (Fleith et al. 2020)

In addition to binder-based and thermal-energy-based techniques, a regolith consolidation process involving only pressure, i.e. potentially less energy-demanding, has been recently demonstrated on Martian regolith simulant. The processes involved rely on the particularities of the Martian environmental conditions and mineralogical composition, which make them specific to this environment (Chow et al, 2017).

### 6.2.4. Conclusions and suggestions concerning future developments

In-situ utilization of local resources, mainly the lunar and Martian regolith, can be envisaged for the production of materials, structures and hardware which are considered essential to address the needs of future long-term exploration activities. A wide range of processes are being investigated, revealing limitations which make them more appropriate to particular uses. Addressing the various infrastructure construction and maintenance needs will likely require a combination of complementary ISRU techniques. For the majority of the investigated processes, a proof of concept has been demonstrated in the terrestrial environment and demonstrator parts have been manufactured. Limited experiments have been conducted to validate the processes in a representative lunar or Martian surface environment. Therefore, further understanding of the effect of the lunar and Martian surface environments on the ISRU manufacturing processes needs to be developed, through relevant testing and analysis. This includes the effect of vacuum, reduced atmosphere and reduced gravity on the regolith processing mechanisms. The impact of the challenging lunar and Martian environments on the processing and manufacturing equipment also needs to be understood, in order to adapt the design of the future manufacturing systems, to ensure their reliability during the missions. In addition, the properties and performance of the manufactured products in their intended environment of application need to be assessed. These development activities will require some level of in-situ testing through demonstration missions, for aspects which cannot be substituted by representative terrestrial laboratory testing.

## 6.3. In-space assembly and manufacturing

### 6.3.1. Introduction

The Moon and Mars are the next destinations for post-International-Space-Station human exploration, as stated in the NASA Global Exploration Roadmap (https://www.nasa.gov/sites/default/files/atoms/files/ger_2018_small_mobile.pdf). This will likely be associated with longer missions in space for the crew, whether it is to conduct scientific investigations(https://exploration.esa.int/web/moon/-/61371-esa-strategy-for-science-at-the-moon

at the Moon), in its orbit or on the surface, and to prepare technologies for exploration to further destinations or to complete the journey to Mars.

From the experience on the International Space Station (ISS), it was recognized that the amount of spares carried out for redundancy purposes, as well as the large quantity of packaging materials represent a major constraint in terms of logistics and space available on the orbital platform 0. An alternative to redundant spares is frequent resupply missions, which also represent significant costs and would not be practical in a long-term journey to Mars.

The ability to manufacture and recycle needed items in space, on-demand, has been identified as key enabler for long duration human exploration missions beyond Low-Earth Orbit. Efforts are being conducted by multiple actors to develop relevant manufacturing technologies, for a range of materials and applications.

Leveraging on the progress in on-orbit manufacturing technologies (Prater, Tracie, et al, 2019), as well as rapid maturation of on-orbit servicing capabilities (https://directory.eoportal.org/web/eoportal/satellite-missions/m/mev-1), technologies for on-orbit manufacturing of large structures in space are being progressed, with the objective of producing spacecraft subsystems which overcome the limitations of launcher fairing size and enable increased performance for potential enhancement of scientific mission outcomes.

## 6.3.2. Requirements from representative missions

On-demand manufacturing capability has been identified as a key element of future long duration human exploration missions (Owens et al, 2016). This would allow to significantly revise mission logistics, by alleviating the need for on-board spares and reducing the dependence on resupply missions. This is particularly relevant for missions where resupply is not practical, such as a human mission to Mars, but also essential to increase the sustainability of long duration human missions in lunar orbit or on the lunar surface. Such capability requires that technologies are available to process items commonly used by the crew or needed for equipment repair and maintenance. This includes parts and equipment made of polymer, metal, ceramics and electronics materials. The ability to recycle materials from discarded items (e.g. defective items, parts at the end of life or packaging material) is also key to sustainable logistics. The in-situ manufacturing of medical equipment or even organs is important to ensure crew safety during the missions.

The outcome of exploration or science missions can be enhanced by increasing the size of certain elements of the spacecraft. This is relevant for solar arrays, which could provide more power – and therefore enable the use of more complex payloads – if they can be made larger. Larger-scale antennas can also lead to increased performance for some applications. The maximum size of these subsystems is currently determined by the size of the available launcher fairing or the capability of deployable systems. On-orbit manufacturing offer the possibility to produce larger subsystems, which could surpass currently available ones.

NASA recently conducted a detailed study to understand when it is worth assembling telescopes in space rather than folding them into a single rocket and deploying them in space (Mukjerjee R.et al, 2019),(Siegler N. et al, 2019). The study concludes, among other findings, that in-space assembly is considered to be a viable approach for observatory assembly and can enable observatory sizes that cannot be achieved by conventional, single-launch approaches.

This requires that relevant manufacturing technologies are available, to operate within the space environment constraints, to produce and assemble the needed structures.

### 6.3.3. Current status and future capacities needed

NASA has been conducting a comprehensive In-Space Manufacturing project since 2014, to develop a range of materials, processes and manufacturing technologies to provide on-demand manufacturing capability for deep space exploration missions (Prater et al, 2019; Clinton Jr et al, 2019). This included the operation of polymer additive manufacturing systems on the ISS, as well as demonstration of polymer recycling on the orbital platform. Additional capability under development in the project include metal additive manufacturing, printed electronics, digital design database and sterilization systems for food and medical applications (Prater et al, 2019).

The Italian Space Agency developed a polymer additive manufacturing system which was operated on the ISS in 2015 with PLA polymer. ESA has since then built an additive manufacturing ground demonstrator system for engineering polymers, demonstrating that it can produce parts, independently of the gravity direction (see Figure 5.23).

(https://www.esa.int/ESA_Multimedia/Images/2018/10/MELT_3D_printer). A metal additive manufacturing system is currently under development, for expected operation on the ISS in 2021 (Figus et al., 2019).

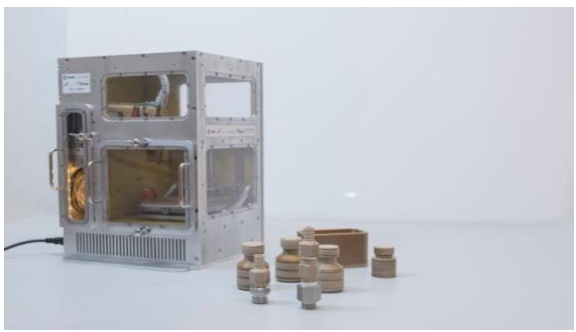

*Figure 5.23. ESA's Manufacturing of Experimental Layer Technology (MELT) additive manufacturing system ground demonstrator, demonstrated to print engineering polymer parts independently of the gravity vector.*

The China Academy of Space Technology is reported to have conducted the first demonstration of on-orbit 3D printing of continuous carbon fibre reinforced polymer composites, during a recent human space flight system test flight (http://www.xinhuanet.com/english/2020-05/09/c_139043414.htm).

Regarding bioprinting capabilities for in-situ medical care, ESA has recently completed a study into technologies for printing of living tissues and the requirements for their applicability in space exploration missions (Cubo-Mateo et al, 2020).

On-orbit manufacturing capability of large spacecraft subsystems, overcoming launcher fairing size limitations, has been the topic of NASA's In-space Robotic Manufacturing and Assembly project 0(Clinton Jr,R.G., et al. 2019). This includes the development of technologies for on-orbit additive manufacturing and assembly of solar array structures and antennae. Manufacturing and assembly of large structures and well as the associated robotic capabilities have already been demonstrated in representative thermal vacuum environments on ground. Demonstration missions are planned in the coming years(COLL et al, 2020), (WILLIAM et al,2020).

In Europe, ESA has led the development of technology for pultrusion of long carbon-fiber-reinforced beams in vacuum, with prospects for application in on-orbit manufacturing of large structures.

(https://www.esa.int/Enabling_Support/Space_Engineering_Technology/Shaping_the_Future/In-Orbit_Manufacturing_of_Very_Long_Booms).

Several internal developments in private entities are also ongoing, to develop relevant on-orbit manufacturing and assembly capabilities (Figus et al, 2019; Patané et al, 2020), which will be relevant for science and exploration mission spacecraft.

### 6.3.4. Conclusions and suggestions concerning future developments

Technologies for in-situ manufacturing during long term human exploration missions have been developed in the last decade, with several of them successfully demonstrated on the ISS. The range of materials for such systems need to be widened, together with multi-material fabrication capability, in order to allow on-demand manufacturing, recycling and maintenance of most items needed during a long exploration mission.

On-orbit manufacturing concepts for large structures and spacecraft subsystems are being matured, through the development of relevant manufacturing processes. Validation of those processes in relevant space environment characteristics need to be continued, until on-orbit demonstration of those technologies.

Such techniques can help enhance the outcome of future exploration and science missions.

## 6.4. Advanced environmental control and life support technology

### 6.4.1. Introduction

Advanced Environmental Control and Life Support (ECLS) will maintain an environment suitable for sustaining human life throughout long duration of human deep space exploration missions. Advanced ECLS technology is very important to the human moon/Mars mission, it concludes the 4 aspects: **Atmosphere Revitalization, Water Recovery and Management, Waste Management, and Environmental Monitoring.** Although ECLS technology has been significantly developed and improved in low-Earth orbit space station missions, it is still a critical technology for long duration missions for human deep space exploration.

## 6.4.2. Requirements from representative missions

For human deep space exploration missions, we can conclude and foresee different representative missions. From the Pillar 2 report in this volume, there are 4 kinds of missions before 2061:

**(1) Return to the Moon:** possible missions include the Artemis project, with first human lunar landings missions completed before 2024, involving no more than 4 crew and no more than 14 days duration, followed by 4-5 times human lunar landing missions before 2030;

**(2) Building a lunar base:** possible missions include building a moon village before 2040, if ISRU technology is mature, which can support more than 50~100 crew for more than a year.

**(3) First human landing on Mars:** possible missions include a first human Mars landing mission before 2040, involving no more than 7 crew and more than 500 days duration.

**(4) Building Mars base:** possible missions include the first human Mars base could be built before 2061.This is clearly along duration enterprise.

**The most important challenge for ECLS systems is longer mission durations with** long duration intermittent uncrewed operational phases. These present challenges to the ECLS system that have not been experienced in previous crewed spacecraft programs.

The ECLS system is considered to be dormant during uncrewed phases since the primary function of life support is not required. However, the dormant ECLS system must still be capable of maintaining the vehicle environment during uncrewed phases in order to sustain remaining vehicle systems and payloads. Additionally, a habitable environment must be reestablished prior to the return of crew, and system health must be maintained to ensure reliable operation through a subsequent crewed phase.

Technologies being developed for exploration missions must consider the long during intermittent dormant operational modes. Sufficient automation must be present to support system maintenance, vehicle contingencies such as a fire event, and fault detection, isolation, and recovery. Further evaluation and testing must be performed on both new and heritage technologies to understand the impacts of long-term dormancy on these systems, and to demonstrate the effectiveness of modifications and operations employed to support intermittent dormancy.

## 6.4.3. Current status and future capacities needed

Whilst the human deep space exploration strategy has increased focus on returning to the lunar surface before embarking on missions to Mars, the overall ECLS system capability evolution objectives remain largely unchanged; namely, to evolve the state-of-the-art ECLS system into a more reliable, more closed-loop system for long duration missions beyond LEO with no resupply from Earth. The overall strategy utilizing the ISS to evolve and test the Exploration ECLS system also remains unchanged; however, where these systems are ultimately deployed in the specific lunar and Mars architecture elements may be slightly different. With the potential for human lunar surface missions occurring prior to Mars transit missions, there may be specific ECLSS functional gaps that will require revision and/or acceleration, such as surface dust filtration and partial-gravity water systems. Table 5.8 shows the ECLS capability gaps between the state of the art and the future mission requirement.

(1) **Atmosphere Revitalization**: advance the atmosphere revitalization (AR) functional area have continued to progress toward exploration mission performance goals. Accomplishments have been realized in the subsystem architecture, oxygen generation and recovery, carbon dioxide removal, and trace contaminant and particulate matter control technical areas.

(2) **Water Recovery and Management**: Although an integrated ECLS system is made up of a variety of subsystems, a major driver in sizing an ECLS system is the Water Recovery Subsystem (WRS). As mission durations increase, recycling water becomes critical. Stored water is inadequate, and wastewater sources must be recycled into potable water. The state of the art WRS used on the ISS relies on a high rate of consumable use (0.032 kg expendables consumed per kg of potable water produced). The urine processor experienced failures due to precipitation caused by the combination of calcium in the urine and sulphuric acid in the urine pre-treatment, and the recovery rate from urine was reduced to approximately 77%. Now that a new pre-treatment formulation using phosphoric acid has been introduced, water recovery can meet or exceed the original 85% design goal. Combined with the percentage of water recovered from humidity condensate, the current overall ISS water recovery rate was 85% but can now reach 93%. For exploration systems the goal established by the Human Health, Life Support, and Habitation Systems Roadmap15 is to reach 98% water loop closure with reduced expendables.

| Subsystem Functional Grouping | Function | Capability Gaps | Orion Short Duration μ-g | Long Duration μ-g | Long Duration Planetary Surface |
|---|---|---|---|---|---|
| Atmosphere Revitalization | CO$_2$ Removal | Improved reliability; ppCO$_2$ <2 mm Hg (2600 ppm) (goal) | | X | X |
| | Trace Contaminant Control | Replace obsolete sorbents with higher capacity; siloxane removal | X | X | X |
| | Particulate Filtration | Surface dust prefilter | | | X |
| | Condensing Heat Exchanger | Durable, chemically-inert water condensation and collection with antimicrobial properties | | X | X |
| | O$_2$ recovery from CO$_2$ | Recover >75% O$_2$ from CO$_2$ | | X | X |
| | O$_2$ generation | Reduced size and complexity, more maintainable | | X | X |
| | High pressure O$_2$ | Replenish 24.8 MPa O$_2$ for EVA; provide contingency medical O$_2$ | | X | X |
| Water Recovery and Management | Disinfection/Microbial Control | Disinfection techniques and technologies for microbial control of water systems, dormancy survival | X | X | X |
| | Wastewater processing | Increased water recovery from urine (>85%), reliability, reduced expendables | | X | X |
| | Urine brine processing | Water recovery from urine brine >90% | | X | X |
| Waste Management | Metabolic solid waste | Low mass, universal waste management system | X | X | X |
| | Non-metabolic solid waste | Volume reduction, stabilization, resource recovery | | X | X |
| Environmental Monitoring | Atmosphere monitoring | Smaller, more reliable major constituent analyzer, in-flight trace gas monitor (no ground samples), targeted gas (event) monitor | X | X | X |
| | Water monitoring | In-flight identification and quantification of species in water | | X | X |
| | Microbial monitoring | Non-culture based in-flight monitor with species identification and Quantification | | X | X |
| | Particulate monitoring | Onboard measurement of particulate hazards | | X | X |
| | Acoustic monitoring | Onboard acoustic monitor | | X | X |

*Table 5.8. ECLS capability gaps.*

Of the various consumables required to sustain human life in space, water accounts for the greatest percentage of material by mass. Spacecraft crews need between 3.5 and 23.4 kg of water per person for each mission day depending on mission requirements. Conversely, spacecraft crews produce between 3.9 and 23.7 kg of waste water per person per day depending on mission requirements. The levels of waste water produced can be higher than water requirements because of contributions from the water content of food and metabolically produced water. The water recovery system on ISS is limited to treating only urine and condensate, which is only about 20% of the potential waste stream on long duration exploration missions, which may include hygiene water, laundry water, and water recovered from brines and solid wastes.

(3) **Waste Management**: Several areas of waste development have made progress over the past year to improve the state of the art. The state of the art in waste management is storage for short periods in flexible soft goods bags for disposal in visiting cargo vehicles. No drying or active door control occurs. Trash storage even for short periods of time in a spacecraft is undesirable and can result in unacceptable aesthetics (odor) or unhygienic conditions (trash escapes, gas evolution, and microbial releases). Fecal waste is collected in rigid metal containers that are stored prior to disposal. Urine is collected, pre-treated and delivered for water recovery. Aboard ISS, the Permanent Multipurpose Module and unused node radial ports are used to temporarily store trash. These volumes are relatively low use habitable volumes so that doors are less noticeable by the crew. During disposal the trash must be manually moved by crew to the departing visiting vehicles through the main habitable volumes. Hence current manually stowed trash methods use valuable habitable volume and crew time as well as contribute to odor and trace contaminant loads. We should begin development of improved metabolic waste collection, trash stowage, and processing technologies.

(4) **Environmental Monitoring**: Environmental monitoring focuses on four elements of the environment in habitable volumes of manned vehicles that have direct impacts on the health and performance of the crew: air quality, water quality, microbial presence, and the acoustic surroundings. Cabin air quality involves monitoring trace Volatile Organic Compounds (VOCs), airborne particles, major constituents, and target gases. Impacting cabin air quality are trace VOCs generated from material outgassing, human metabolic processes, chronic leaks from systems and/or payloads, as well as visiting vehicles. Major constituents include oxygen, nitrogen, carbon dioxide, water vapour, hydrogen, methane, and argon. Monitoring trace VOCs require a broader dynamic mass with high sensitivity due to the wide variety of VOCs observed, typically in the sub-parts per million concentration range. Major constituent monitoring requires a relatively narrow dynamic mass range but a broad concentration range from a few parts per million to percentages.

Target gases are a subgroup of VOCs that typically require monitor portability to be able to measure in different locations within the manned vehicle. Pockets of higher-than-normal concentrations of a target gas can form due to the lack of convection in microgravity coupled with non-ideal air circulation. This situation arises during maintenance behind a rack or when circulation is disrupted due to temporary stowage of hardware and/or supplies in open areas. Water quality involves the determination of water potability and simultaneously serves as a useful monitor of the system health of water processing hardware. This capability involves the ability to identify and quantify aqueous species which is currently performed on the ground with return samples: an impractical solution for long duration, Exploration-class missions. Aqueous species in the water come from several sources including any wetted material associated with the water processing system as well as water-soluble VOCs in the cabin air. Pertinent to water monitoring is the monitoring of biocide levels to ensure appropriate levels are maintained for the health of the crew.

(5) Microbial monitoring identifies and quantifies the microbial presence aboard manned vehicles. Similar to water quality, effort focuses on addressing the lack of return samples and ground analysis. Consequently, work focuses on technology that can identify microbial presence and potentially quantify as well.

(6) Real-time acoustic monitoring ensures the noise levels in the cabin are below the limits set forth by the NASA Acoustics Office. The resulting acoustic environment in an enclosed volume arises from the sum of all the noise sources. As such the acoustic levels will vary according to what is operating. As the cabin environment approaches the acoustic limits for long-term auditory health, crossing that limit can be difficult to discern. Crew may be exposed to greater acoustic levels than perceived for a significant amount of time without any indication. This can impact not only long-term auditory health, but it may also impact performance and communication.

# 6.4.4. Conclusions and suggestions concerning future developments

Various country and agency efforts have endeavoured to define and refine concepts for human exploration beyond LEO. Throughout this, the ECLS capability development efforts have remained relevant and continue to make steady progress. Additional considerations associated with architectural evolution, and intermittent dormancy are being taken into consideration in the various technology efforts.

# 7. Disruptive technologies

New technologies being developed under the "NewSpace" umbrella are being rolled out at a pace that has the potential to disrupt planetary science exploration over the next four decades. Rates of infusion for new technologies have dropped from decades in some cases to just a few short years. Rates are dropping so fast that, by 2061, small spacecraft with mass 50-200 kg may be able to do what today's 500-1000 kg spacecraft can achieve. It will be common practice to incorporate CubeSat/NanoSat ride-alongs on flagship missions to enable science measurements at close range and in environments that would be considered too risky for the primary spacecraft. Small landers will allow us to explore the surface and even the subsurface of planetary bodies. Science results from these smaller, subsidiary missions will often have a higher profile than results from the primary mission, and attract much greater public attention – as Philae did on Rosetta. Recent trends also suggest launch costs/kg will be at an all-time low and capabilities at an all-time high.

Telecom, always challenging for deep space missions to the outer planets, will benefit from downlink rates using optical communications that will match today's rates for inner planet missions using RF. The rapid advances made in communication rates from CubeSats in Low Earth Orbit (LEO) over the last decade are illustrative of the pace of infusion. In 2012, the fastest downlink from a CubeSat to a ground station was 1200 bps using UHF communication channels. Eight years later the record is held by Aerospace Corporation's Optical Communications and Sensors Demonstration (OCSD) CubeSat (Rose et al, 2019) which used an optical communication link to demonstrate 100 Mbps, with higher rates (600 Mbps) expected soon. Similar rates were demonstrated with NASA/JPL's ISARA mission, which has an innovative reflectarray coating applied to the backplane of its deployable solar panels to create a Ka-band high gain antenna, capable of up to 100 Mbps downlink (Hodges et al, 2013). The same reflectarray technology also enabled the success of MarCO's X-band downlink in 2018 – a mere 8 kbps – but that was from Mars (Foust, 2018). Thus, in less than eight years, the state-of-the-art in CubeSat telecom went from amateur levels of capability to advanced levels in LEO, and a demonstration of useful telecom rates on the first interplanetary CubeSats.

What can we expect for telecom on future planetary missions? NASA's Psyche mission will carry the Deep Space Optical Communication out to the asteroid belt and demonstrate optical communication from as far out as 3.3 AU from the Sun (https://www.nasa.gov/mission_pages/tdm/dsoc/index.html). Assuming the trend towards outer planets missions using solar power continues, as seen in NASA's Juno and Europa Clipper, and ESA's JUICE mission, a promising technology that would allow the large-area solar arrays to double-up as High Gain Antennas can be found in microwave reflective, optically transparent coatings (Saberin, 2010). It's also reasonable to expect that Information bandwidth will have increased dramatically as onboard science data reduction in deep space missions becomes commonplace, lowering the need for ever-higher downlink rates. Space-qualified data processing capability on deep space missions (currently strangled by the 1990's era Rad750 with clock rates of order 100 MHz) will be just a few years behind the ground-based processing capability of 2050, which will be blindingly fast. In fact, the state-of-the-art for deep space missions may soon see a huge boost due to the successful deployment of a QualComm Snapdragon 801 processor on the Ingenuity helicopter currently operating at Mars, with a 2.26 GHz clock rate (Grip et al, 2019). With much faster processor speeds, software functionality (AI, autonomy, fault protection, data processing and analytics) on-board spacecraft will grow rapidly from the present state.

In the future, hardware upgrades for long-lived spacecraft in Earth orbit using additive manufacturing technology or satellite servicing will be as common as uploading software upgrades is today. We should expect that additive manufacturing will be used successfully in a low-gravity environment to construct large-scale structures, e.g. a habitat, or a space telescope or a very large antenna. Substantial progress is already being made in this direction, as seen by the award of a NASA Contract worth $73.7M to Made In Space for its Archinaut project, with the aim of constructing a pair of 10-meter solar arrays in LEO (https://arstechnica.com/science/2019/07/nasas-technology-program-funds-ambitious-in-space-manufacturing-mission).

Spacecraft structures will be multifunction without exception – providing structural integrity, thermal conduction, comm lines, power distribution, and even RF/optical reflecting surfaces (Aglietti et al, 2007). All spacecraft subsystems and instrument components will be the result of detailed simulation and be 3-D printed (Stevenson, 2018). Integration and test will be almost fully automated, as foreshadowed by the incredible production rate of 2 satellites per day in OneWeb's Florida facility [10]. *The formulation/design phase will still take the 2-3 years it does now – but fabrication, integration and test will shorten to a time-span of just a few weeks, even for one-off spacecraft like those we send out to explore deep space.*

[1] Solar cell efficiencies may have reached a plateau by 2061, and batteries should be available that operate efficiently in all expected temperature regimes for deep space missions, from Venus to as far out as Pluto and beyond. We will have already seen an electromagnetic tether prototype generate power, first in LEO （https://phys.org/news/2018-01-power-propulsion-satellites.html）.but by then on at least one outer planets mission. Propulsion using electromagnetic tethers that exploit the Lorentz force by interacting with Earth's ionosphere will also soon be a reality (https://www.scientificamerican.com/article/kilometer-long-space-tether-tests-fuel-free-propulsion/). For deep space missions, the same principle will enable magnetoshell aerocapture to shave off 100's of kgs of fuel required to get into orbit around Mars and the outer planets (https://www.nasa.gov/feature/magnetoshell-aerocapture-for-manned-missions-and-planetary-deep-space-orbiters/). Advances in power and propulsion technologies based on fission or even fusion nuclear processes beyond present-day capabilities could be available, depending in part on the direction Earth-based nuclear power systems take over the next four decades.

Attitude determination and control systems will continue their advance towards sub-arcsec pointing control, as demonstrated by the '2018 SmallSat of the Year' ASTERIA （Smith et al, 2018; and steady progress will continue towards cm-level precision in formation flying (http://utias-sfl.net/?p=2191, to the point where these are no longer considered a risk item. Science remote sensing instruments will continue to shrink in power requirements and physical size (Freeman et al, 2020), with the exception of measurements requiring large apertures, like Synthetic Aperture Radars. In those cases, the mass of the structure forming the aperture will continue to trend downwards (Freeman et al. 2018).

Advances in mobility systems will have made the static landers of today obsolete. Drones, crawlers and penetrators will be engaged in a detailed geologic survey of Mars conducted from the surface, and will have made extensive explorations of subsurface lava tubes (https://www.space.com/alien-life-hunt-mars-underground.html). We will have explored similar structures on the Moon that could allow planetary geologists to study the history of ancient lava eruptions, and serve as radiation-protected habitats for future human explorers (https://www.smithsonianmag.com/science-nature/nasa-considers-rover-mission-go-cave-diving-moon-180971790/). Fueling stations will provide access to top-ups of H2 and O2 generated in-situ, so they can continue to explore for years at a time. With the development of MOXIE, a pathfinder payload on NASA's Mars 2020 rover, currently en route to Mars, we may soon see in-situ fuel production for the first time on another planet (https://mars.nasa.gov/mars2020/spacecraft/instruments/moxie/). Aerobots drifting in Venus' atmosphere will raise and lower their altitude to look for further signs of life following up on the recent exciting report of elevated levels of phosphine in the atmosphere (Greaves et al, 2020). Venus aerobots may also provide glimpses of active surface processes beneath the clouds (Babu et al, 2020). The first cryobot explorers will be swimming in the oceans of Europa and Enceladus (https://www.space.com/alien-life-ocean-moons-europa-enceladus.html). Drones will have given us a close-up view of the complex hydrocarbon chemistry in Titan's atmosphere(https://en.wikipedia.org/wiki/Dragonfly_(spacecraft)).

1. Over the past decade, there have been some incredible developments in SmallSat technology, with more to come in the near future (Malphrus et al,2020). The first launch of the SLS rocket, expected soon, will propel 13 CubeSats into Deep Space, most headed for the Moon. ESA's Hera mission to the Didymos binary asteroid pair will carry two ride-along CubeSats to be deployed on arrival, as will the complementary NASA mission to the same target, DART. But this technology does not have to be confined to SmallSats: a recent white paper submitted to the National Academy in the US for the Planetary Science and Astrobiology Decadal Survey looked at the use of SmallSat technology to reduce costs and enhance capabilities for a mission to one of the Ice Giants in the outer solar system (http://surveygizmoresponseuploads.s3.amazonaws.com/fileuploads/623127/5489366/24-99acda7716fc45b6b1b3029e61192c89_BalintTiborS1.pdf). These projected developments, taken together, mean that, despite the "tyranny of the rocket equation," we can expect planetary science missions in 2061 to go further and faster than they do now, touch more objects in our solar system, return far more information, and be implemented on budgets and schedules we can only dream of today.

# 8. Conclusions

It would not be possible to elaborate in this already overlong article all the technologies needed, or likely to be available, for planetary exploration over the next 40 years. Areas of technology which must be addressed certainly include:

- New Rockets/Launch systems (including solar sails, elevators)
- New in-space propulsion systems, notably high thrust electric/plasmic propulsion
- Power: Nuclear generators (RTG etc), Fuel Cells, Advanced solar cells and batteries, low power electronics.
- Communication via radio or laser links, or other techniques?
- Machine Learning
- Autonomy and on-board data reduction
- Navigation – 'Fire-n-Forget' navigation - tell us when you get there
- Low cost systems (incl. cubesats, penetrators)
- Advanced fabrication techniques, both in terrestrial production and in-situ
- Large planetary telescopes…
- Sample Return and Curation
- Improved Simulations
- Laboratory experiments

The instrumentation for planetary missions continues to evolve at speed, as new opportunities, missions and science questions emerge. Directions for future instrumentation will include:

- Life detection – non-DNA specific detection approaches
- Miniaturisation - Instrument on a chip
- Multipoint in situ measurements
- New Detector technologies
- New techniques for composition measurements, including isotopes.
- Contamination control –
- Sample return – cryogenic and hot capability
- Autonomy to optimize science data collection

Our aim has been rather to highlight key areas, and illustrate by example the types of innovations which will be required, and what we can expect. Some bottlenecks are very obvious, such as the need for new power technologies for long duration missions, or the need for a balance between sample return and more sophisticated in-situ and autonomous measurement. In some cases this will require long and expensive developments.

In other cases, terrestrial technologies, for example in DNA sequencing and life detection, machine learning, or autonomous vehicles, which are all currently developing at break-neck speed, will spin off into planetary exploration with entirely new and disruptive consequences.

It is hoped that the topics covered in this article can provide insight into what those developments will be, and a basis for a roadmap for effective planning of future technology research and development for planetary exploration.

*Acknowledgements: Part of the work described in section 7 was carried out by the Jet Propulsion Laboratory, California Institute of Technology, under a contract with the National Aeronautics and Space Administration. VS would like to thank Ralf Srama for his careful reading and helpful suggestions concerning the dust instrumentation section.*